\documentclass[pre,amsmath,amssymb,aps,twocolumn,superscriptaddress,longbibliography,floatfix]{revtex4-2}
\usepackage{amssymb}
\usepackage{amsmath}
\usepackage{commath}
\usepackage{verbatim}
\usepackage{graphicx}
\usepackage[english]{babel}
\usepackage{dcolumn}
\usepackage{bm}
\usepackage[dvipsnames]{xcolor}
\usepackage[colorlinks]{hyperref}
\usepackage{graphicx}
\usepackage{array}
\usepackage{physics}
\newcolumntype{o}{@{}>{{}}c<{{}}@{}}
\hypersetup{                
	colorlinks=true,                
	breaklinks=true,                
	urlcolor= blue,                
	linkcolor= red,                
    bookmarksopen=false,
	filecolor=black,
	citecolor=blue,
	linkbordercolor=black
}

\graphicspath{ {images/} }

\begin{document}
\title{Mirror transitions in diffusion with stochastic resetting confined on a ring}
\author{Pedro Juli\'an-Salgado}
 \email{pjuliansalgado@gmail.com}
\affiliation{Departamento de Física, Universidad Autónoma Metropolitana, Mexico City 09340, Mexico}%
\author{Pavel Castro-Villarreal}%
 \email{pcastrov@unach.mx}
\affiliation{Facultad de Ciencias en F\'isica y Matem\'aticas, Universidad Aut\'onoma de Chiapas, Carretera Emiliano Zapata, Km. 8, Rancho San Francisco, 29050 Tuxtla Guti\'errez, Chiapas, M\'exico.}%
\author{Leonardo Dagdug}%
 \email{dll@xanum.uam.mx}
\affiliation{Departamento de Física, Universidad Autónoma Metropolitana, Mexico City 09340, Mexico}%
\author{Denis Boyer}
 \email{boyer@fisica.unam.mx}
\affiliation{Instituto de Física, Universidad Nacional Autónoma de México, Mexico City 04510, Mexico}

\begin{abstract}
    Diffusion with an incorporated resetting mechanism provides a reference framework for modeling a wide range of natural phenomena. Within this framework, the optimal resetting rate is a key quantity that arises from the optimization of the mean first-passage time. While substantial work has focused on the study of the optimal resetting rate in unbounded one dimensional domains, little is still known about the optimization of the mean first-passage time in bounded systems, in particular 
    when multiple resetting sites are available. In this work, we consider a particle diffusing along a circular circumference and under resetting, with an absorbing target site at a fixed location. Using the appropriate free propagator for this system, we compute the Laplace transform of the survival probability when resetting occurs to multiple sites drawn from an arbitrary probability density function. We also calculate the mean first-passage time at the target site, and study the dependence of the optimal resetting rate in terms of the relevant parameters of the system in a two-resetting site configuration. Depending on the arc length between one of the resetting sites and the absorbing target site, and the weight of the remaining resetting site, the optimal resetting rate can exhibit abrupt (``first order'') and continuous (``second order'') transitions. Moreover, the behavior of the mean first-passage time is rich enough to allow both critical and tri-critical points to exist in the parameter space. All the transitions have ``mirror symmetry'' around the selected target site and its corresponding diametrically opposite site.
\end{abstract}

\maketitle
\section{Introduction}

Searching agents, either microscopic or macroscopic, often engage in dynamic strategies to locate specific sites or resources essential for sustaining biological processes. At the microscopic level, for example, specialized proteins or transcription factors diffuse in their environment and bind to specific DNA sequences \cite{mechetin_mechanisms_2014,tafvizi_dancing_2011,kolomeisky_physics_2011,halford_how_2004,kamagata_single-molecule_2023,berg_diffusion-driven_1981}. Living organisms such as animals actively forage in their environment in search of food and reproductive partners \cite{1990search,louapre_male_2015}. Frequently, these movements occur under time and energy constraints, and can become efficient through optimization. According to the optimal foraging theory, individuals that adopt strategies maximizing net energy gain while minimizing their costs--such as time and effort--are more likely to survive and reproduce \cite{king_optimal_2022,cowie_optimal_1977,CHARNOV1976129}. L\'evy-type processes, for instance, have been incorporated into optimal foraging theory, and have motivated models to explain the scale-free movements observed in empirical data of large animals \cite{bartumeus_levy_2011,atkinson_scale-free_2002,boyer_scale-free_2006,viswanathan_optimizing_1999,ares_modeling_2010,viswanathan_physics_2011}. The inactivation of diffusive molecular agents after a finite search time also finds applications in cellular and intracellular transport \cite{ma_strong_2020,lawley_effects_2021,meerson_mortality_2015}. Works related to activated transport cell mechanisms, which are constrained to finite search times, have also been developed and include systems that range from immune migrating cells searching for harmful pathogens \cite{shaebani_distinct_2022,krummel_t_2016}, optimal run and tumble dynamics \cite{rupprecht_optimal_2016} or molecular motors such as kinesins and dyneins \cite{hirokawa_kinesin_2009}.

In most of the situations mentioned above, the search processes to be optimized take place in environments where inherent stochastic dynamics strongly influence first-passage encounters \cite{Condamin_confined_subdiffusion}. Brownian particles in unbounded domains, on average, take an infinite time to arrive at a fixed region of space or site. One simple strategy to render this time finite is to suddenly stop the ``search phase'', and execute a faster phase of relocation in which the objective target remains non-detectable by the search agent. Such intermittent search models have proven to be optimizable through the tuning of parameters controlling the relocation and search times, and are widespread, as they commonly describe a variety of systems across different scales \cite{benichou_intermittent_2011,kramer_behavioral_2001}. 

Diffusion with a resetting protocol can be classified as an intermittent search process. In these systems, stochastic motion represents a phase during which the target may be captured, whereas a resetting event acts as a fast relocation that drives the particle back to a pre-established state. Remarkably, this relocation action eliminates inefficient trajectories that would otherwise cause the mean first-passage time (MFPT) to diverge in unbounded domains. This feature was studied by Evans \& Majumdar in their seminal work \cite{evans_diffusion_2011}, where they also showed that diffusion under stochastic resetting leads to the emergence of non-equilibrium stationary states (NESS)--a topic subsequently developed in other works \cite{majumdar_dynamical_2015,Eule_2016,pal_diffusion_2015}. Diffusion subject to stochastic resetting has attracted considerable attention and has been studied for many systems. Subsequent extensions and generalizations of the original formulation have further enriched the theoretical foundations of this field \cite{evans2014diffusion,pal_diffusion_2016,evans_diffusion_2011-1,pal2019first,Mercado-Vasquez_2020,chechkin2018random,evans2018effects,Radice_2022}. We can mention random walks on networks with resetting to a node \cite{riascos2020random}, the thermodynamics of stochastic resetting \cite{fuchs2016stochastic}, threshold resetting \cite{2020optimization,biswas2025target}, or experimental realizations \cite{besga_optimal_2020,faisant_optimal_2021,friedman_resetting_experiment}, to name just a few. Applications leveraging the intermittent dynamics of stochastic processes with resetting are diverse and span several scientific areas. Models using this framework are suitable, for instance, in the description of systems susceptible to catastrophic events \cite{Asymmetric_Plata_Carlos}, the emergence of correlations in systems composed of independent particles  \cite{extreme_Biroli_2023}, strategies to prevent global-warming-related abnormal events in climate models \cite{ali_asymmetric_2022}, frameworks examining climate-related decisions at long-horizon via stochastic models of discounting \cite{montero2022valuing}, predator-prey systems with site relocation \cite{mercado2018lotka,evans2022exactly}, or cell division \cite{genthon2022branching} among others.

Going beyond the standard framework of resetting to a single site or state, a growing body of work has demonstrated that allowing multiple resetting locations can modify first passage optimization in a non-trivial way \cite{gonzalez_diffusive_2021}. In one dimension, resetting to two discrete sites already produces a remarkably rich structure: the MFPT may develop two competing minima, and the optimal resetting rate can undergo discontinuous transitions as the relative position or statistical weights of the resetting sites are varied \cite{two_resetting_points}.

This phenomenology also occurs in the general case where the resetting position is chosen randomly from a probability density function \cite{besga_optimal_2020,faisant_optimal_2021,abrupt_2025}, where the MFPT can also exhibit abrupt transitions and critical lines separating smooth and discontinuous regimes. Works describing transitions in the MFPT based on a Landau-like framework have also been developed \cite{pal2019landau}. In diffusion in external potentials, the optimal resetting rate may vanish either continuously or discontinuously depending on the potential barrier asymmetry and boundary conditions \cite{ahmad2022first}. Furthermore, a general framework interpreting resetting transitions in terms of the interaction between interacting thermal and potential energy scales has been proposed in \cite{ray_resetting_2021}. In geometrically constrained systems, such as diffusion in narrow channels described by the Fick-Jacobs approach, resetting modifies first passage properties in a non-trivial way and can lead to multiple extrema in the MFPT \cite{jain2023fick}. Another example is provided by Brownian search under a fixed time constraint in which resetting events incur an explicit cost, so that the optimal strategy follows from balancing resource expenditure against the probability of successfully locating the target \cite{sunil2024minimizing}.

L\'evy flights under resetting exhibit discontinuous changes in the parameters minimizing the MFPT \cite{kusmierz2014first}, while restart protocols in quantum walk dynamics lead to multiple minima in mean detection times \cite{yin2023restart}. Optimization induced by a resetting mechanism has also been discussed in reaction-kinetic settings of the type Michaelis-Menten, where the rate of unbinding (resetting) can switch from being beneficial to detrimental depending on systems parameters \cite{rotbart2015michaelis,pal2019landau}.

These works show that when resetting takes place in the presence of spatial heterogeneity, multiple return sites, external potentials, or non-Gaussian transport, the MFPT is likely to develop a complex dependence on the resetting rate, leading to non-trivial optimization characterized by the coexistence of several local minima. However, while stochastic resetting has been previously explored for searchers on circular geometries \cite{chatterjee2018}, the interplay between a curved manifold and the optimization of multi-site resetting remains unexplored. Standard Brownian motion on curved manifolds has been a subject of interest for several decades. For example, the foundational work by A. Kolmogoroff introduced a covariant formulation of the Fokker-Planck (FP) equation \cite{kolmogoroff_zur_1937}, which was subsequently studied by Stratonovich \cite{stratonovich_auxiliary_1992}, who identified the role of Riemannian geometry through the diffusion tensor. More recently, M. Polettini \cite{polettini_generally_2013}, employing a gauge principle in a covariant Langevin equation, derived the covariant formulation of the FP equation, a result also obtained via holonomic constraints \cite{castro-villarreal_brownian_2014}. The covariant formulation of stochastic equations has seen significant development over the past decade (e.g., see \cite{giordano_stochastic_2019,polettini_generally_2013,castro-villarreal_covariant_2023}). Earlier studies have also focused on understanding the behavior of quantities such as the mean-square displacement as a function of the manifold geometry. It is generally recognized that short-time behavior is influenced by curvature properties of Riemannian geometry, while long-time behavior is governed by global invariants of the Riemannian metric; see the recent review \cite{solano-cabrera_self-assembly_2025} and references therein for a discussion of Brownian motion on curved spaces. The present work examines Brownian motion with a resetting protocol to multiple sites on the simplest non-trivial one-dimensional manifold, the circle $S^{1}$.

This work is organized as follows: section \ref{general_framework} reviews the general framework of the renewal approach for the survival probability and mean first-passage time under stochastic resetting to random positions. This framework is then applied in section \ref{one_res_point} to a Brownian particle on a circle of radius $R$ with a single resetting site to derive a closed-form expression for the mean first-passage time and identify the regimes where resetting expedites the search process given a fixed target location, extending previous work on diffusion under resetting in an interval \cite{pal2019first}. In section \ref{sec:results}, we consider the same problem with two resetting sites, a configuration previously studied on the semi-infinite line and known to admit a rich phenomenology in first-passage observables \cite{two_resetting_points,abrupt_2025}. Elucidating its effects within the manifold $S^{1}$ is a primary focus of this work. Furthermore, in this section, we compute the averaged mean first-passage time (AMFPT) for the maximum arc configuration between the primary resetting site and the target. We identify tri-critical points that separate continuous and discontinuous transitions in the optimal resetting rate, occurring as the weight of the second resetting site is varied. Subsequently, in section \ref{other_segments} we analyze the order parameter diagram for alternative circular segments by shifting the primary resetting angle closer to the target. We conclude this section with Table \ref{table}, which summarizes the critical and tri-critical coordinates for other representative configurations between the primary resetting site and the target location. Finally, in section \ref{conclusions}, we provide our concluding remarks.

\section{General framework}\label{general_framework}
In this section, we review a general framework for studying first-passage statistics of an arbitrary process with stochastic resetting to successive positions $z_1,z_2,\ldots$, that are i.i.d. random variables taken from a probability density function (PDF) $\rho(z)$ \cite{evans_diffusion_2011-1,two_resetting_points,abrupt_2025}. In order to accomplish this task, we use a renewal equation for the survival probability in the presence resetting. We define $Q_r(x_0,t)$ as the probability that the particle under resetting and with initial position at $x_0$ has not yet found a specified target region up to time $t$. Hereafter, unless explicitly indicated, the term target denotes an absorbing site, meaning that once the particle reaches it for the first time, it exits the system. For the sake of notation, we omit the explicit dependence of $Q_r(x_0,t)$ on the distribution $\rho(z)$.

The dynamics for a process with stochastic resetting with rate $r$ is as follows: during an infinitesimal interval $[t,t+\dd  t]$, with probability $r\dd t$ the particle is instantaneously relocated to a position $z$ drawn from $\rho(z)$. With the complementary probability $1-r\dd t$, it continues its underlying process. Since reset events constitute a Poisson process, the intervals between successive resets are independent and exponentially distributed. With these considerations we write the renewal equation for the survival probability as \cite{evans_diffusion_2011-1,evans_stochastic_2020}
\begin{multline}
	Q_{r}\left( x_{0},t\right) =e^{-rt}Q_{0}\left( x_{0},t\right)\\
	+r\int_{-\infty}^{\infty}\rho(z) \dd z\int_{0}^{t}e^{-r\tau }Q_{r}\left( x_{0},t-\tau \right) Q_{0}\left(
	z,\tau \right) \dd\tau.  \label{gen_sur}
\end{multline}
Where $Q_0(x_0,t)$ is the survival probability of the underlying process without resetting with initial condition $x_0$. For example, for a free one-dimensional Brownian particle in a semi-infinite domain with an absorbing target at the origin, this probability is given by $Q_{0}(x_0,t)={\rm erf}\big(\frac{|x_{0}|}{\sqrt{4Dt}}\big)$. 

The expression for $Q_r(x_0,t)$ in Eq. \eqref{gen_sur} is made up of two terms. The first term on the right-hand side (rhs) represents trajectories in which no reset event occurred up to time $t$, which occur with probability $e^{-rt}$, thus the particle survives with probability $Q_0(x_0,t)$. The last term on the rhs is a double average over $z$ and $\tau$, due to the contribution of surviving trajectories that underwent their last reset event from a position $z$ at a time $t-\tau$. Hence the particle has not found the target with probability $Q_r(x_0,t-\tau)$ until the last resetting, and has not found it either during its resetting-free path of duration $\tau$.

From first-passage theory we know the MFPT (here denoted as $T_r(x_0)$) of an arbitrary process, is related to the survival probability through \cite{dagdug_diffusion_2024}
\begin{equation}
    T_r(x_0) = \int_0^{\infty}Q_r(x_0,t) \dd t \label{mfpt_def}.
\end{equation}
Let us define the Laplace transform of $Q_r(x_0,t)$ as $\tilde{Q}_{r}\left( x_{0},s\right) =\mathcal{L}\{Q_r(x_0, t)\}=\int_{0}^{\infty
}e^{-st}Q_{r}\left( x_{0},t\right) \dd t$. Clearly, setting $s=0$ gives $\tilde{Q}_{r}\left( x_{0},0\right) =\int_{0}^{\infty
}Q_{r}\left( x_{0},t\right) \dd t$. Hence, the MFPT above is related to the Laplace transform of the survival probability through
\begin{equation}
 T_r(x_0)=\tilde{Q}_{r}\left( x_{0},s=0\right).   \label{lapla_surv_mfpt}
\end{equation}
Taking the Laplace transform of Eq. \eqref{gen_sur} yields
\begin{equation}
	\tilde{Q}_{r}\left(x
    _0,s\right) =\frac{\tilde{Q}_{0}\left(
		x_0,r+s\right) }{1-r\int_{-\infty}^{\infty}\rho(z)\tilde{Q}_{0}\left(z,r+s\right) \dd z },  \label{laplace_surv}
\end{equation}
where $\tilde{Q}_{0}\left( x_{0},s\right) =\mathcal{L}\{Q_0(x_0, t)\}$. From Eq. \eqref{lapla_surv_mfpt}, we deduce
\begin{equation}
	 T_r\left( x_{0}\right)=\tilde{Q}
	_{r}\left( x_{0},s=0\right) =\frac{\tilde{Q}_{0}\left(
		x_0,r\right) }{1-r\int_{-\infty}^{\infty}\rho(z)\tilde{Q}_{0}\left( z,r\right) \dd z }.  \label{mfpt}
\end{equation}
Consequently, the MFPT with an incorporated resetting mechanism can be computed using the Laplace transform of the survival probability of the underlying process without resetting. 

In the following sections, we will be interested in the optimization of the MFPT, i.e., we will study the values of $r$ that make the MFPT of Eq. \eqref{mfpt} a minimum. The optimal resetting rate $r^{\ast}$ satisfies
\begin{equation}
r^{*}(x_0) = \underset{r\ge 0}{\arg \min}\,T_r(x_0). \label{rast_def}
\end{equation} 

The effect of a small resetting rate in a given system is beneficial for search if the MFPT is a decreasing function of $r$ in the limit $r\to 0$, or

\begin{equation}
	\lim_{r\rightarrow 0}\frac{\dd T_{r}}{\dd r}<0.  \label{cond_res}
\end{equation}
In the case of resetting to a single point $x_1$, where $\rho(z)=\delta(z-x_1)$,
for resetting to expedite the search process, the following condition must be fulfilled (see \cite{reuveni_optimal_2016,pal2017first} and also appendix \ref{ap:cv}), 
\begin{equation}
	\langle t_{0}^{2} (x_0) \rangle
	-2 \langle t_{0}(x_0)\rangle
	\langle t_{0}(x_1)\rangle>0,  \label{cond_res1}
\end{equation}%
where the random variable $t_{0}(x_i)$ is the first-passage time of the underlying
process without resetting, starting from the position $x_i$.

\section{Optimization of the MFPT with one resetting point on \texorpdfstring{$S_1$}{}}\label{one_res_point}
We now apply the general framework of the previous section to the following system with periodic boundary conditions: a Brownian particle confined to a circle of constant radius $R$.
%%%%%%%%%%%% Figure 1 %%%%%%%%%%%%%%%
\begin{figure}
\centering
\includegraphics[width=\linewidth]{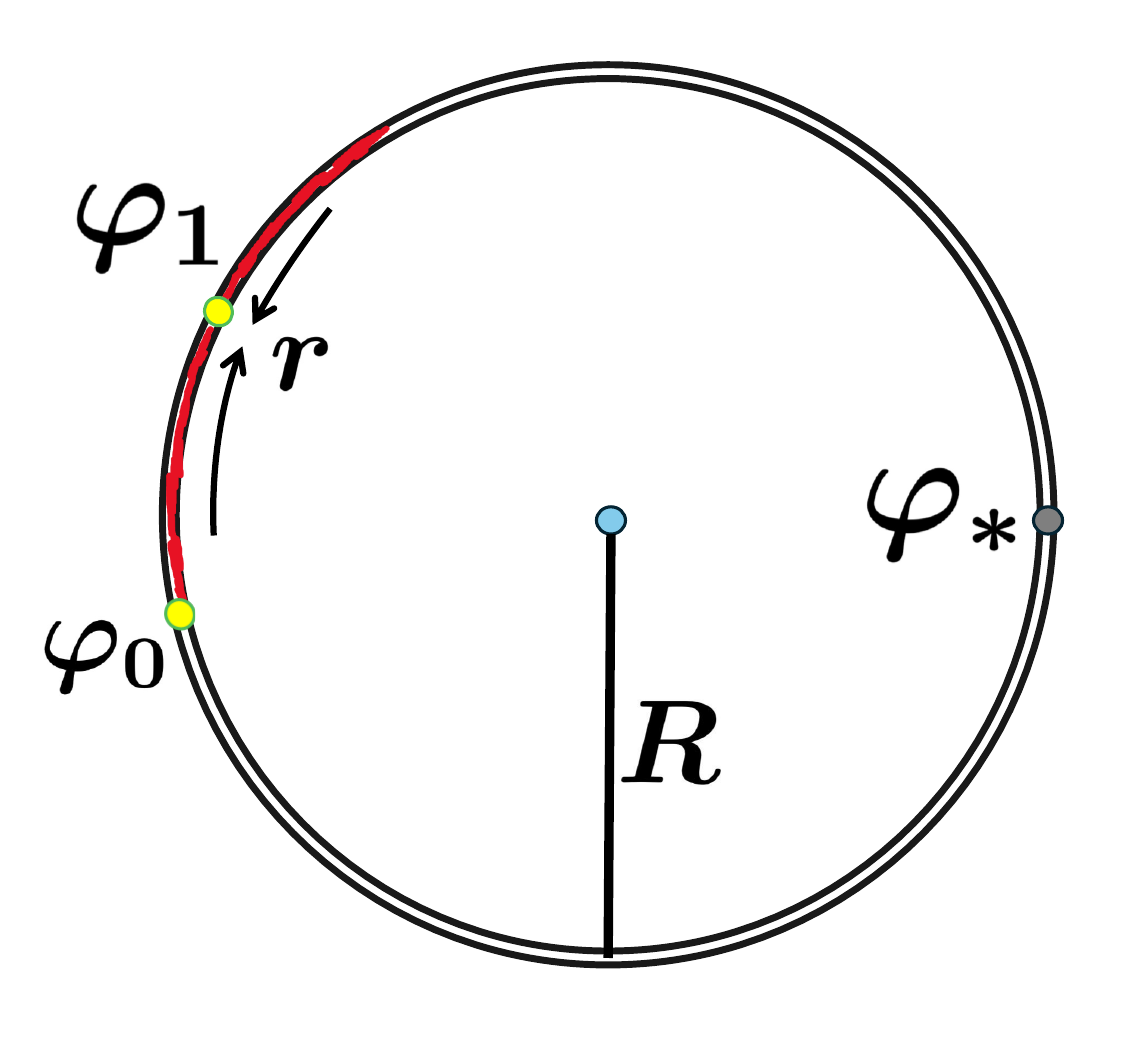}
  \caption{One resetting site configuration for a Brownian particle in a ring of radius $R$. The particle starts its path from $\varphi_0$, and diffuse along the circumference (a trajectory is shown in red). At random times with rate $r$ the particle resets to $\varphi_1$. The process ends when it first reaches the absorbing point at $\varphi_\ast$.}
	\label{fig:illustration1}
\end{figure}
%%%%%%%%%%%%%% End of figure 1 %%%%%%
Let the particle's position be described by the angular coordinate $\varphi(t)\in[0,2\pi]$, which plays a role analogous to the spatial variable $x(t)$ in the interval. The particle initiates motion at $\varphi_0$ and undergoes normal diffusion along the circle. We introduce a Poissonian resetting mechanism with constant rate $r$, by which the particle is instantaneously relocated to a fixed angular position $\varphi_1$. We distinguish the resetting position from the initial one, i.e., $\varphi_0 \neq \varphi_1$ in general. The process ends when the particle reaches for the first time a stationary target located at $\varphi_{\ast}$, as depicted in Fig. \ref{fig:illustration1}.

In order to use the MFPT result derived in Eq. \eqref{mfpt}, we identify the analogous quantities for this specific geometry. In the general framework, resetting occurrs to a position $z$ drawn from a distribution $\rho(z)$. For the ring scenario with a single, fixed resetting site (considered in this section), $z$ is an angle $\theta$ and the distribution is a Dirac delta function centered at $\varphi_1$:
\begin{equation}
     \rho(\theta)=\delta(\theta-\varphi_1) .\label{first_dist}
\end{equation}
By substituting $\rho(\theta)$ into the denominator of expression \eqref{mfpt} we obtain
\begin{equation}
    T_r(\varphi_0) = \frac{\tilde{Q}_0(\varphi_0,r)}{1-r\tilde{Q}_0(\varphi_1,r)}.\label{ring_mfpt}
\end{equation}
Here, $\tilde{Q}_0(\varphi_0,s)$ represents the Laplace transform of $Q_0(\varphi_0,t)$, the probability that a free Brownian particle starting at $\varphi_0$ has not yet visited the target $\varphi_{\ast}$ by time $t$. The Laplace transform $\tilde{Q}_0(\varphi_0,s)$ can be calculated by means of the memoryless property of diffusion and using standard methods for solving partial differential equations (e.g. separation of variables). The full result for $\tilde{Q}_{0}\left( \varphi_{0},s\right)$ is calculated explicitly in appendix \ref{surv_eq}, and is given by 
\begin{equation}
    \tilde{Q}_{0}\left(\varphi _{0},r\right) =\frac{1}{r}\left[ 1-\frac{%
		\cosh \xi_{r}\left( \pi -\left\vert \varphi _{\ast }-\varphi
		_{0}\right\vert \right) }{\cosh \xi_{r}\pi }\right], \label{surv_expl}
\end{equation}
where $\xi_r$ is a dimensionless parameter that depends on the resetting rate $r$ through
\begin{equation}
    \xi_r = R\sqrt{\frac{r}{D}}, \label{xi_expl}
\end{equation}
where $D$ is the diffusion coefficient and $R$ the radius of the ring. In addition, we recall that $\varphi_0$ and $\varphi_\ast$ denote the initial and absorbing angular positions, respectively.

Substituting Eq. \eqref{surv_expl} into Eq. \eqref{laplace_surv} yields a closed-form expression for $\tilde{Q}_r(\varphi_0,s)$
\begin{equation}
	\tilde{Q}_{r}\left( s,\varphi _{0}\right) =\frac{\cosh \xi _{l}\pi -\cosh
		\xi _{l}\left( \pi -\left\vert \varphi _{\ast }-\varphi _{0}\right\vert
		\right) }{s\cosh \xi _{l}\pi +r\cosh \xi _{l}\left( \pi -\left\vert \varphi
		_{\ast }-\varphi _{1}\right\vert \right) },  \label{closed_one_sur}
\end{equation}
with $\xi _{l}=R\sqrt{(r+s)/D}$. Furthermore, according to \eqref{ring_mfpt} and \eqref{surv_expl}, the MFPT is
\begin{equation}
	T_{r}\left( \varphi _{0}\right) =\frac{\cosh \xi_r \pi -\cosh \xi_r \left( \pi -\left\vert \varphi _{\ast }-\varphi _{0}\right\vert \right) 
	}{r\cosh \xi_r \left( \pi -\left\vert \varphi _{\ast }-\varphi
		_{1}\right\vert \right) }.
	\label{exp_mfpt}
\end{equation}
This is the main result for this section. The first takeaway from this equation is the finiteness of $T_r$ when $r\to0$:
\begin{equation}
	T_{0}\left( \varphi _{0}\right) =\frac{R^{2}}{2D}\left( \pi ^{2}-\left( \pi
	-|\varphi_{\ast}-\varphi_{0}|\right) ^{2}\right).  \label{t0_eq}
\end{equation}
On the other hand, one can corroborate that in the limit $r\to \infty$, Eq. \eqref{exp_mfpt} diverges. Indeed, in this limit the particle hardly explores its surroundings in search of the target, as it is reset to the resetting site too frequently.

\begin{figure}
\centering
\includegraphics[width=\linewidth]{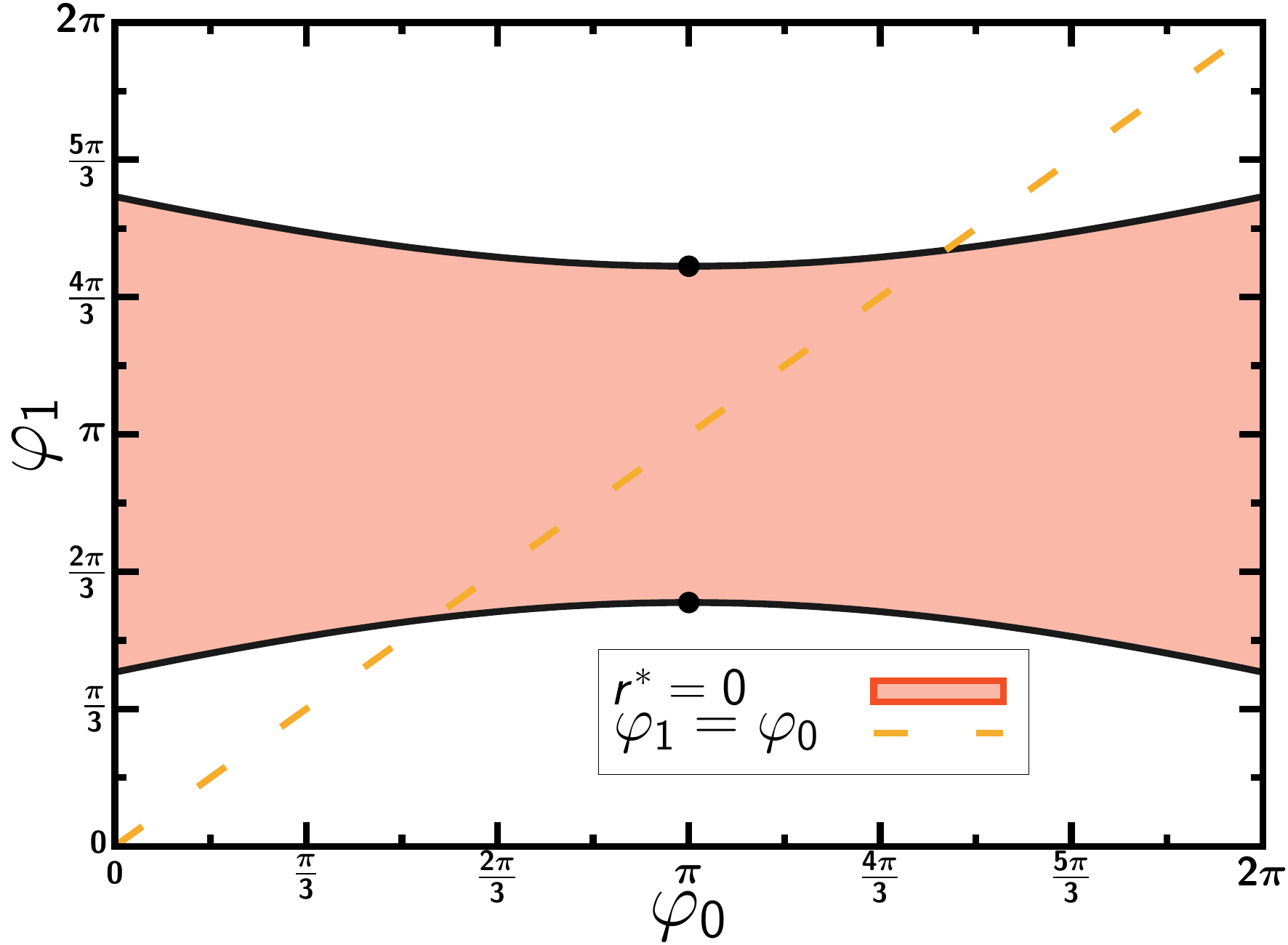}
  \caption{In white, we represent the pair values $(\protect\varphi_0,\protect%
		\varphi_1)$ at which resetting can expedite the MFPT of Eq.
		\eqref{exp_mfpt}. The dashed line corresponds to the special case when $\varphi_0 = \varphi_1$.}
	\label{fig:dia_phir_vs_phi0}
\end{figure}

Furthermore, we can also explore the regimes where resetting expedites first passage encounters between the Brownian searcher and the target. According to Eq. \eqref{cond_res1}, in order to accomplish this task we need to calculate the second moment, $\left \langle t_0^{2}(\varphi_i)\right \rangle$, of the underlying process without resetting related to the set-up shown in Fig. \ref{fig:illustration1}. According to Eq. \eqref{ap_b3b} of appendix \ref{ap:cv}, the second moment of any process can be written as
\begin{equation*}
   \langle t_{0}^{2}(\varphi_{i})\rangle =2\int_{0}^{\infty
	}dt\, tQ_0(\varphi_{i},t). 
\end{equation*}
This last equation can also be rewritten
in terms of $\tilde{Q}_0$
\begin{equation}
  \langle t_{0}^{2}(\varphi_{i})\rangle=-2  \lim _{s\to0}\dv{\tilde{Q}_0(\varphi_i,s)}{s} .\label{lp_drivative}
\end{equation}
Hence, we first expand $\tilde{Q}_0(\varphi_i,s)$ of Eq. \eqref{surv_expl} in powers of $s$
\begin{multline}
    \tilde{Q}_0(\varphi_i,s) = \frac{R^{2}}{2D}\left( \pi^{2}-a^{2}\right)\\ +\frac{R^{4}}{4D^{2}}\left(\pi^{2}a^{2} - \frac{5\pi^{4}}{6} -\frac{a^{4}}{6}\right)s +\mathcal{O}(s^{2})\label{expansion_lt_Q}
\end{multline}
where $a = \pi - |\varphi_{\ast}-\varphi_i|$. Using this expansion, along with Eq. \eqref{expansion_lt_Q}, it is straightforward to show that the second moment of the MFPT in Eq. \eqref{t0_eq}, denoted as $\left \langle T^{2}_0(\varphi_i) \right \rangle $, is
\begin{equation}
    \left \langle T_{0}^{2}(\varphi_i)\right \rangle = \frac{R^{4}}{2D^{2}}\left(-\pi^{2}|\varphi_i - \varphi_{\ast}|^{2} + \frac{5\pi^{4}}{6} +\frac{|\varphi_i - \varphi_{\ast}|^{4}}{6}\right) .\label{second_moment}
\end{equation}
Substituting Eqs. \eqref{t0_eq}-\eqref{second_moment} into the condition \eqref{cond_res1}, and upon a simplification we obtain
\begin{multline}
\frac{1}{24}\big( -\varphi
	_{0}^{4}+6\varphi _{0}^{2}\varphi _{1}^{2}+4\pi \varphi _{0}^{3}-12\varphi
	_{0}^{2}\varphi _{1}\pi\\ -12\pi \varphi _{0}\varphi _{1}^{2}+24\pi
	^{2}\varphi _{0}\varphi _{1}-8\varphi _{0}\pi ^{3}\big) >0,
	\label{s1-13B3}
\end{multline}%
where we have reduced the number of free parameters by taking the target location at $\varphi_\ast = 0$ without restricting generality. The inequality in Eq. \eqref{s1-13B3} is satisfied if $\varphi _{1}\in U$, where
\begin{equation}
    U=\left[ 0,u_{-}\right] \cup \left[
u_{+},2\pi \right] \label{u_interval}
\end{equation}
and $u_-(\varphi_0)$, $u_+(\varphi_0)$ are given by
\begin{equation}
\begin{split}
	u_{-}(\varphi_0) =\frac{6\pi -\sqrt{6}\sqrt{\varphi _{0}^{2}-2\varphi _{0}\pi +2\pi
			^{2}}}{6} \\
	u_{+}(\varphi_0) =\frac{6\pi +\sqrt{6}\sqrt{\varphi _{0}^{2}-2\varphi _{0}\pi +2\pi
			^{2}}}{6}. \label{limits}
\end{split}%
\end{equation}
The restart mechanism is beneficial for the task completion
if $\varphi _{1}$ is sufficiently close to the target site at $\varphi_*=0$. Here, $u_-$ serves as an upper bound for $\varphi_1$ on $[0,\pi]$, whereas $u_+$ provides a lower bound for $\varphi_1$ on $[\pi,2\pi]$. This information can be visualized in Fig. \ref{fig:dia_phir_vs_phi0}, where we show the configuration space plot $(\varphi_0,\varphi_{1})$, along with the corresponding bounds $u_-(\varphi_0)$ and $u_{+}(\varphi_0)$ for $\varphi_1$. For this and all subsequent figures in this work, unless otherwise specified, the ring radius and diffusion coefficient are set to unity, i.e., $R=1$ and $D=1$. However, note that the results for Fig. \ref{fig:dia_phir_vs_phi0} are independent of $R$ and $D$.
%%%%%%%%%%%% Figure 2 %%%%%%%%%%%%%%%
%%%%%%%%%%%%%% End of figure 2 %%%%%%

Our diagram in Fig. \ref{fig:dia_phir_vs_phi0} is divided into three zones, two of them (in white) represent the set of values $(\varphi_0,\varphi_1)$ that make the resetting
mechanism optimum for a non-zero value of $r^{\ast}$.  Meanwhile, the red zone highlights values of $%
\varphi_0$ and $\varphi_1$ at which it is better not to use resetting, or $r^{\ast} =0$. The white region is the largest when $\varphi_0=\pi$, where
\begin{eqnarray}
  u_-(\varphi_0 = \pi) &=&  u^{(max)}_- = \frac{6\pi -\sqrt{6}\pi}{6}\label{max}\\
 u_+(\varphi_0 = \pi) &=& u^{(min)}_+ = \frac{6\pi+\sqrt{6}\pi}{6}  .\label{min}
\end{eqnarray}
Hence, choosing $\varphi_0 =\pi$ minimizes the region where resetting does not help the particle to find the target. 
The complementary region, which we call $U^{\text{max}}$, is explicitly given by Eqs. \eqref{max}-\eqref{min} 
\begin{equation}
    U^{\text{(max)}} = [0,u_-^{\text{(max)}}]\cup[u_{+}^{\text{(min)}},2\pi].
\end{equation}
The dashed line of Fig. \ref{fig:dia_phir_vs_phi0} corresponds to the case where the resetting point is the initial condition, $\varphi_1 = \varphi_0$ in Eq. %
\eqref{s1-13B3}. Resetting is favorable in this case if $\varphi_0 \in U_0$ with
\begin{equation}
U_0=[0,\frac{(5-\sqrt{5})\pi}{5}] \cup [\frac{ (5 + 
	\sqrt{5})\pi }{5},2\pi].
\end{equation}
This interval is contained in Fig. %
\ref{fig:dia_phir_vs_phi0} as it is characterized by the intersections of the dashed line
with the black curves $u_{-}$ and $u_{+}$ at: $u_{-} \left( \frac{(5-\sqrt{5}
	)\pi}{5} \right) = \frac{(5-\sqrt{5})\pi}{5}$, $u_{+} \left( \frac{(5+\sqrt{%
		5 })\pi}{5} \right) = \frac{(5+\sqrt{5})\pi}{5}$. Observing that $U^{\text{(max)}}$ exceeds  $U_0$ in length, it follows that allowing $\varphi_1 \neq \varphi_0$ broadens the region in which resetting enhances the average first-passage performance.
        Pal and Prasad in \cite{pal2019first} obtained analogous results for a one dimensional interval $[a,b]$ with absorbing boundaries at both ends, when the initial position is identical to the resetting one.

%%%%%%%%%%%% Figure 3 %%%%%%%%%%%%%%%
\begin{figure}
\centering
\includegraphics[width=\linewidth]{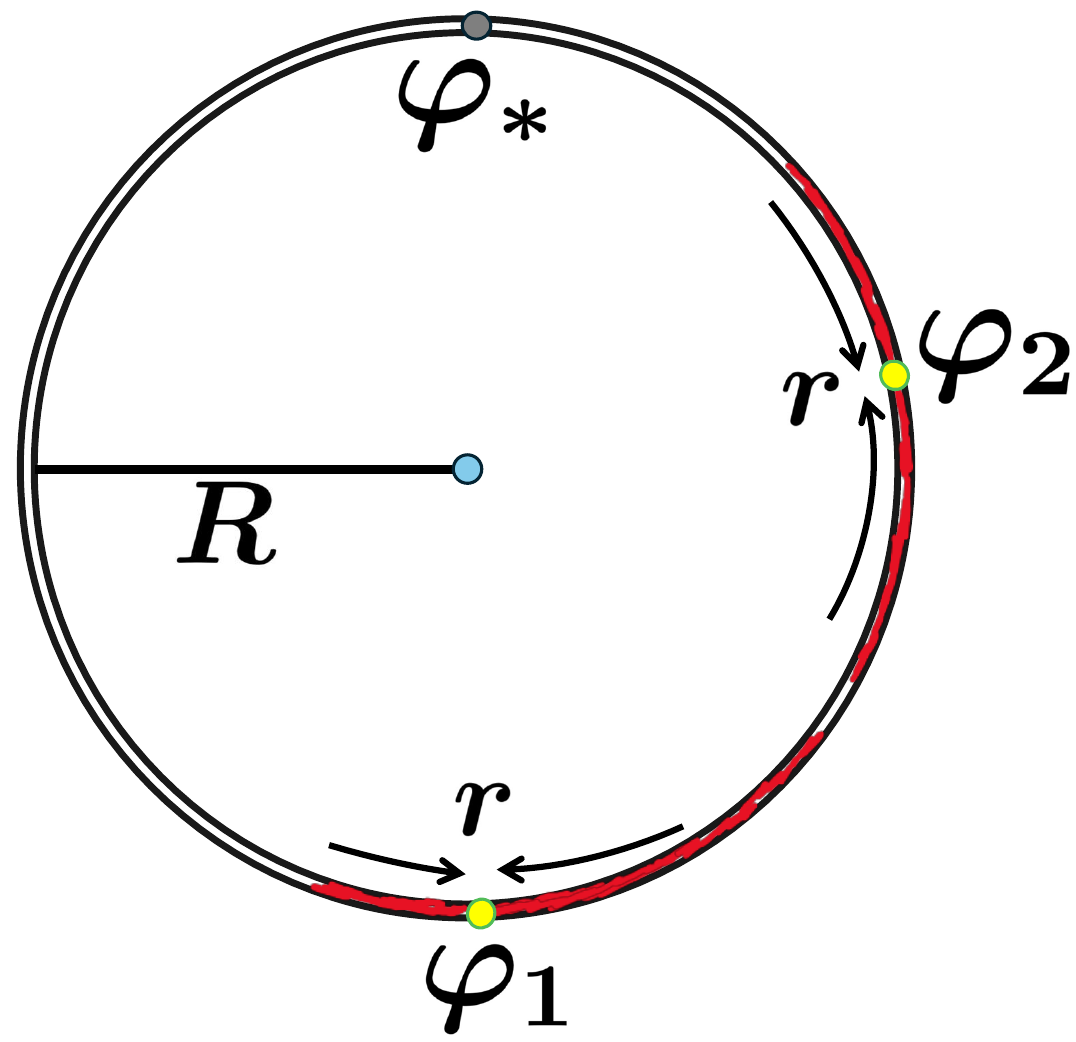}
  \caption{System configuration for a Brownian particle having two distinct angular sites to reset in a ring: $\varphi_1$ and $\varphi_2$. At random times, with rate $r$ the particle resets to $\varphi_1$ with probability $1/(1+m)$, or to $\varphi_2$ with the complementary probability $m/(1+m)$. In this set-up we considered the absorbing target $\varphi_{\ast}$ and $\varphi_1$ to be diametrically opposed by setting $\varphi_1 = 3\pi/2$ and $\varphi_{\ast}=\pi/2$.}
	\label{fig:illustration2}
\end{figure}
%%%%%%%%%%%%%% End of figure 3 %%%%%%

\section{Resetting to multiple sites: the averaged mean first passage time}\label{multiple_sites}
The above analysis can be extended to a richer and more complex scenario: a Brownian particle with access to multiple resetting sites. To address this case, we recall that Eq. \eqref{mfpt} provides a general expression for the MFPT which can be applied to a diffusion process with stochastic resetting, valid whenever the instantaneous relocation site is drawn from an arbitrary PDF $\rho$. This observation allows us to directly adapt the theoretical framework reviewed in section \ref{general_framework} to the ring geometry. In particular, we now interpret $\rho(\theta)$ as the PDF that governs the probability density of the possible angular resetting sites. With these considerations, we now write Eq. \eqref{mfpt} as
\begin{equation}
	 T_r\left( \varphi_{0}\right)=\frac{\tilde{Q}_{0}\left(
		\varphi_{0},r\right) }{1-r\int_{0}^{2\pi}\rho(\theta)\tilde{Q}_{0}\left( \theta,r\right) \dd \theta }.  \label{mfpt_phi}
\end{equation}
Anticipating the multi-parameter structure of the MFPT, we now consider the initial angle $\varphi_0$ to be also distributed according to $\rho(\theta)$. This assumption removes one explicit parameter from the problem and allows us to obtain a more compact expression for the MFPT. Thus, we define the averaged mean first passage time (AMFPT) as
\begin{equation}
    \left \langle T_{r} \right \rangle = \int_{0 }^{2\pi }\rho
		\left( \theta\right) T_r(\theta)\dd \theta= \frac{\int_{0 }^{2\pi }\rho
		\left( \theta\right) \tilde{Q}_{0}\left( \theta,r\right) \dd \theta}{1-r\int_{0}^{2\pi }\rho \left( \theta\right) \tilde{Q}_{0}\left(\theta,r\right) \dd \theta}. \label{averaged_mfpt}
\end{equation}
Here, the AMFPT is denoted with angular brackets to distinguish from the MFPT. Hence, the Brownian particle begins its trajectory from an initial angle selected randomly among all possible resetting sites, according to the PDF $\rho(\theta)$. Substituting Eq. \eqref{surv_expl} for Brownian diffusion on the circle into Eq. \eqref{averaged_mfpt}  gives
\begin{equation}
	\left \langle T_{r} \right \rangle=\frac{1}{r}\left[ \frac{\cosh \xi_r\pi 
	}{\int_{0 }^{2\pi}\rho \left( \theta\right) \cosh \xi_r\left( \pi
		-| \varphi _{\ast }-\theta| \right) \dd \theta}-1\right],  \label{s1-13C}
\end{equation}
where $\xi_r$ is given by Eq. \eqref{xi_expl}, and, as in the previous section, $\varphi_\ast$ is the target site location. 

In what follows, we exploit this rather simple expression to investigate the optimization behavior of $\langle T_r\rangle$ for only two resetting sites.

The PDF for two resetting sites, denoted by $\varphi _{1}$ and $\varphi _{2}$ in Fig. \ref{fig:illustration2}, can be written as
\begin{equation}
\rho \left( \theta\right) =\frac{1}{m+1}\delta \left( \theta-\varphi
_{1}\right) +\frac{m}{m+1}\delta \left( \theta-\varphi _{2}\right), \label{two_res_site}
\end{equation}
where $\delta(x)$ is the Dirac delta, $m$ a non-negative number and $p_1 = 1/(1+m)$, $p_2=m/(1+m)$ are the probabilities to reset to $\varphi_1$ and $\varphi_2$, respectively. In this sense, $m$ is a parameter that controls the resetting weight of the site $\varphi_2$. Substituting Eq. \eqref{two_res_site} into \eqref{s1-13C} we obtain
\begin{widetext}
\begin{equation}
	\left\langle T_{r}\right\rangle =\frac{1}{r}\left[ \frac{\left( m+1\right)
		\cosh \xi_r\pi }{\cosh \xi_r\left( \pi -\left\vert \varphi _{\ast
		}-\varphi _{1}\right\vert \right) +m\cosh \xi_r\left( \pi -\left\vert
		\varphi _{\ast }-\varphi _{2}\right\vert \right) }\right]\\
        -\frac{1}{r}.  \label{s1-13D}
\end{equation}
\end{widetext}

\section{Results on two-site resetting with $\varphi_{\ast}=\pi/2$ and $\varphi_1=3\pi/2$ }\label{sec:results}

To characterize the optimal behavior of the AMFPT in Eq. \eqref{s1-13D}, we can reduce the number of free parameters by fixing $\varphi_{\ast}$ and $\varphi_1$. In this section, we start by focusing on the case where these angles are diametrically opposite on the circle (see Fig. \ref{fig:illustration2}).  We examine in particular the variations  of the optimal rate $r^{\ast}$ that minimizes the AMFPT as a function of $\varphi_2$ and $m$. In Section \ref{other_segments} we will consider several other values of $\varphi_1$ chosen progressively closer to the target location $\varphi_{\ast}$.

\subsection{Analysis of the AMFPT}

Let us fix the target site at $\varphi_{\ast}=\pi/2$ in all cases and study the AMFPT in Eq. \eqref{s1-13D} for the specific choice $\varphi
_{1}=\frac{3\pi }{2}$, allowing $\varphi _{2}$ to vary over the entire interval $\left[ 0,2\pi \right] $. Implementing these values into Eq. \eqref{s1-13D} we have%
\begin{equation}
	\left\langle T_{r} \right\rangle =\frac{1}{r}%
	\left[ \frac{\left( m+1\right) \cosh \xi_r \pi }{1+m\cosh \xi_r\left(
		\pi -\left\vert \frac{\pi }{2}-\varphi _{2}\right\vert \right) }-1\right],  \label{s1-13E}
\end{equation}
%
%%%%%%%%%%%% Figure 4 %%%%%%%%%%%%%%%
\begin{figure}
\centering
\includegraphics[width=\linewidth]{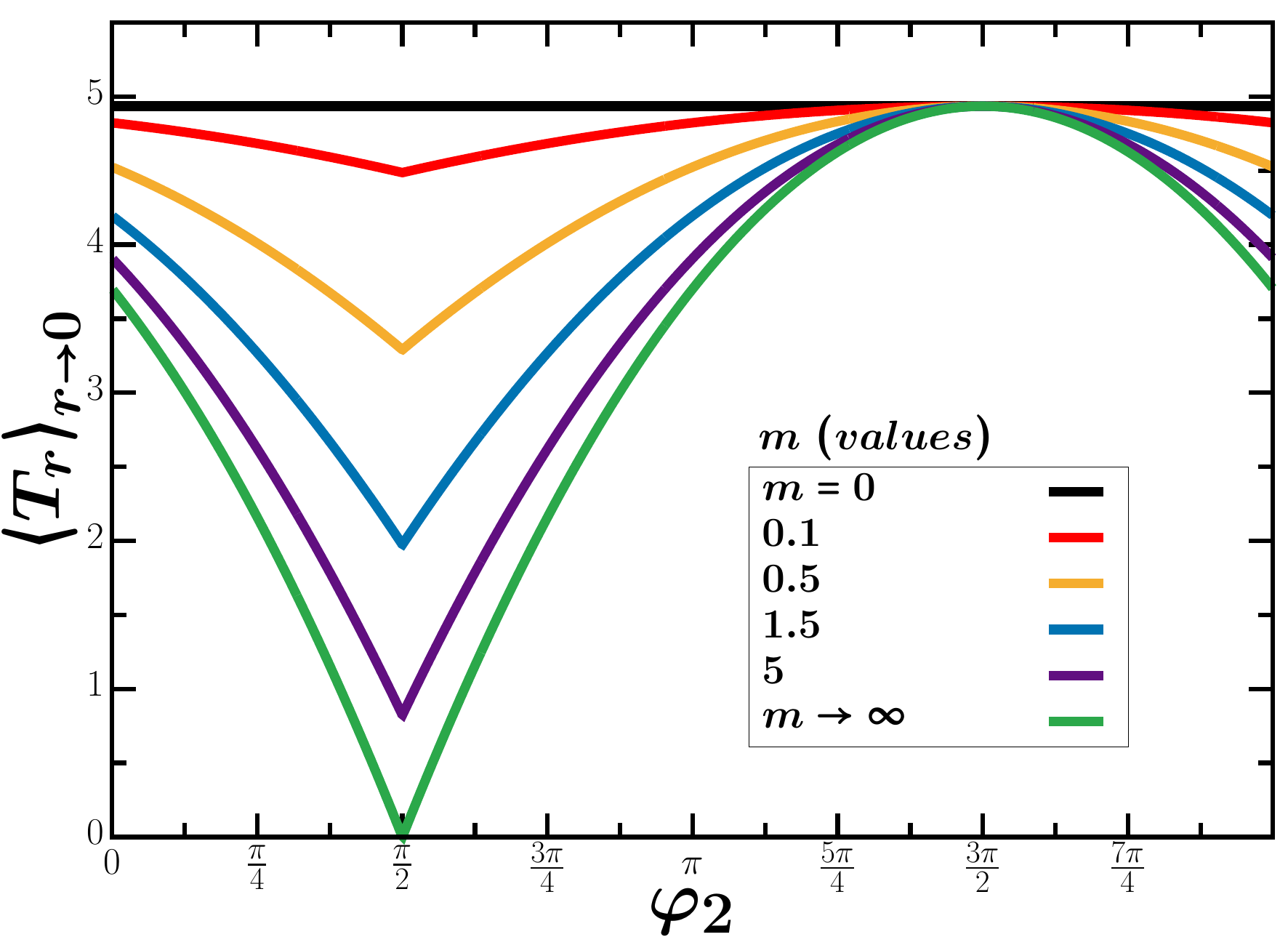}
  \caption{AMFPT in Eq. \eqref{s1-15}. In this no-resetting limit $\varphi_1$ and $\varphi_2$ are the possible initial sites of the free Brownian particle. The target and the first resetting site are fixed at $\varphi_{\ast}=\pi/2$ and $\varphi_1 = 3\pi/2$, respectively.}
	\label{fig:t0_vs_phir}
\end{figure}
%%%%%%%%%%%%%% End of figure 4 %%%%%%
%
where $\xi_r$ is given by Eq. (\ref{xi_expl}). Relevant limits of this expression are now summarized:

\begin{itemize}
\item When $m=0$ this expression corresponds to the MFPT with
one resetting point in equation \eqref{exp_mfpt}, with $\varphi _{0}=\varphi
_{1}=\frac{3\pi }{2}$ and $\varphi _{\ast }=\frac{\pi }{2}$, or%
\begin{equation*}
	\left\langle T_{r}\left( r\right) \right\rangle =\frac{1}{r}\left[ \cosh \xi_r\pi -1\right] .
\end{equation*}%

\item When $m\rightarrow \infty $ (the second resetting point at $%
\varphi _{2}$ is the only one where resetting occurs) the expression in %
\eqref{s1-13E} has the limit 
\begin{equation}
	\lim_{m\rightarrow \infty }\left\langle T_{r}
	\right\rangle =\frac{1}{r}\left[ \frac{\cosh \xi _{r}\pi }{\cosh \xi
		_{r}\left( \pi -\left\vert \frac{\pi }{2}-\varphi _{2}\right\vert \right) }-1%
	\right] ,  \label{s1-14}
\end{equation}%
which is equation \eqref{exp_mfpt} with $\varphi _{0}=\varphi_1=\varphi_{2}$ and $\varphi _{\ast }=\pi/2$. 

\item The limit $r\rightarrow 0$ of Eq. \eqref{s1-13E} is%
\begin{equation}
	\lim_{r\rightarrow 0}\left\langle T_{r}
	\right\rangle =\frac{R^{2}\pi ^{2}}{2D}-\frac{mR^{2}\left( \pi -\left\vert 
		\frac{\pi }{2}-\varphi _{2}\right\vert \right) ^{2}}{2D\left( 1+m\right) }.
	\label{s1-15}
\end{equation}%

The dependence of Eq. \eqref{s1-15} on the resetting angle $\varphi_2$ is shown in Fig. \ref{fig:t0_vs_phir}, where curves corresponding to different values of $m$ are presented. The two limits, when $m=0$ and $m\rightarrow \infty $, are
represented by the black and green curve, respectively. Analytically, those
limits are $\lim_{r\rightarrow 0}\left\langle T_{r}\left( r,\varphi
_{2},m=0\right) \right\rangle =R^{2}\pi ^{2}/2D$ and $\lim_{r\rightarrow
	0}\left\langle T_{r}\left( r,\varphi_{2},m\rightarrow \infty \right)
\right\rangle =R^{2}\left( \pi ^{2}-\left( \pi -\left\vert \frac{\pi }{2}%
-\varphi_{2}\right\vert \right) ^{2}\right) /2D$. Clearly, the only
case where the AMFPT is zero without resetting is when the searcher starts its path at $\pi /2$ with probability one.
\end{itemize}

\begin{figure}
  \centering
    \includegraphics[width=0.95\linewidth]{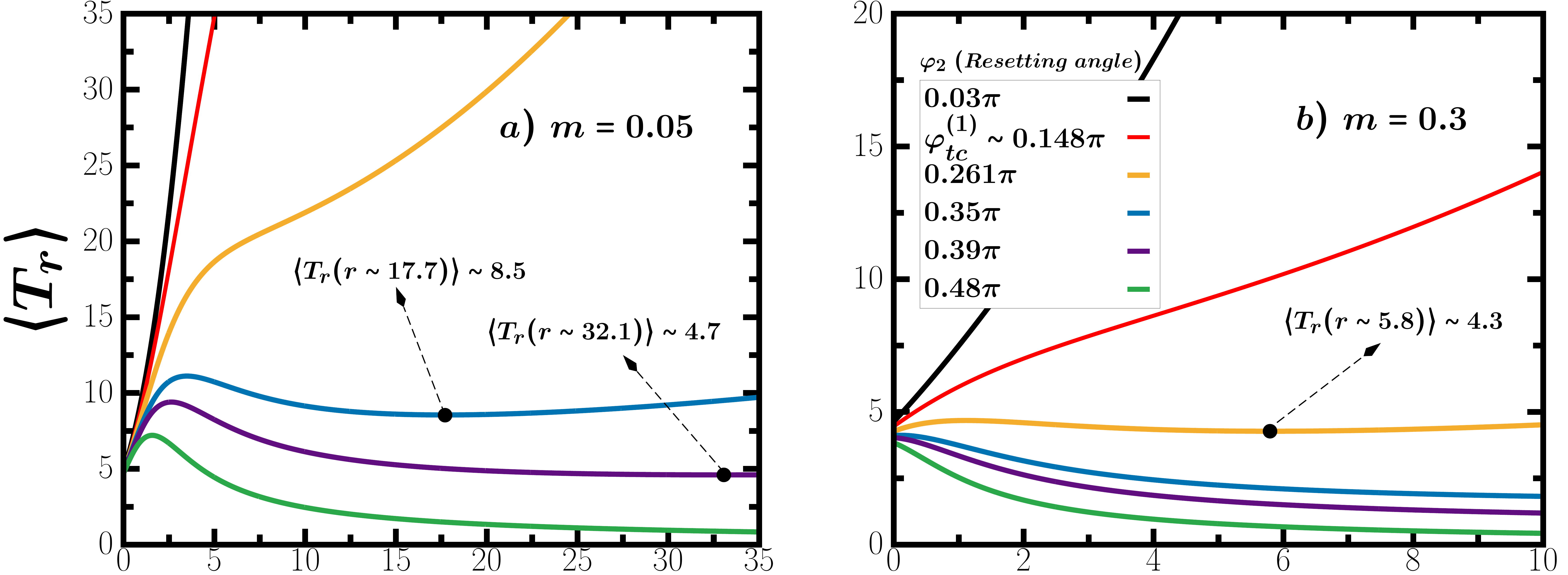}
    \label{fig:t1_a} \\
    \includegraphics[width=0.95\linewidth]{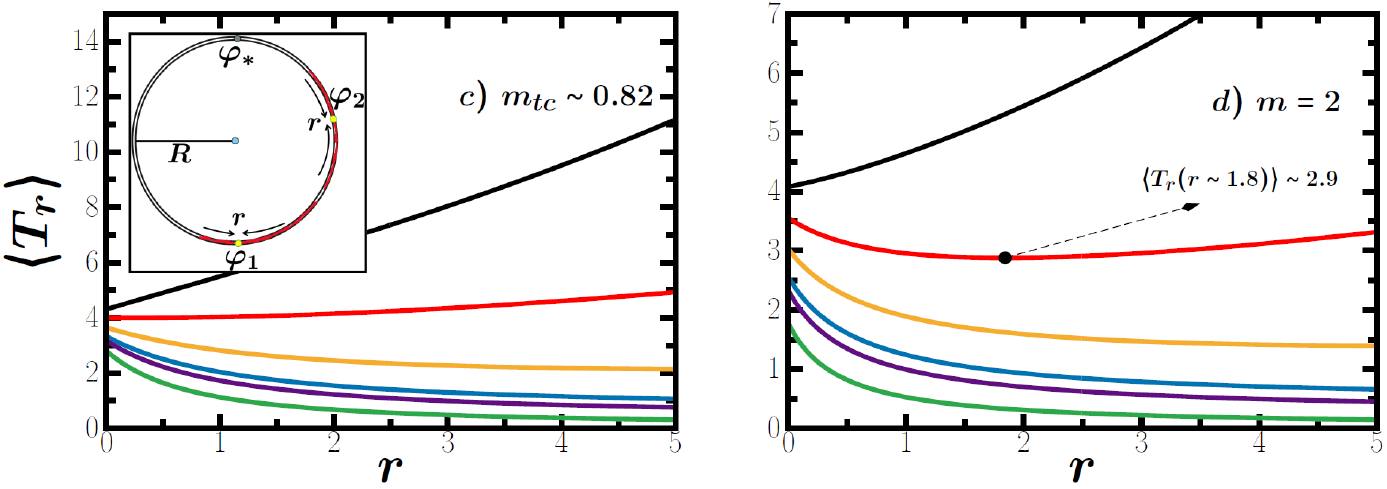}
    \label{fig:t2_b}
  \caption{AMFPT as a function of $r$. Each panel represents a fixed value of $m$ and each curve within a panel, a fixed value of $%
		\protect\varphi_{2}$ (same color code). All $%
		\protect\varphi_{2}$ values belong to the range $\protect\varphi_{2} \in [0, 
		\protect\pi/2]$. For clarity, the inset of panel c) recalls the system under consideration, also shown in Fig. \ref{fig:illustration2}.}
	\label{fig:t_vs_r_0-1d57}
\end{figure}

\noindent Intricate behaviors emerge for finite values of $m$ and $r$. In Fig. \ref{fig:t_vs_r_0-1d57}, we display $%
\left\langle T_{r}\left( r,\varphi_{2},m\right) \right\rangle $ as a
function of the resetting rate $r$. Each panel corresponds to a representative value of $m$ and six curves show the behavior of %
Eq. \eqref{s1-13E} for specific values of $\varphi_{2}\in [ 0,\pi /2]$ (the same in all panels). 

\noindent When a local minimum occurs at a non-zero $r$ within
the plotted range, it is marked with a black dot. 

In the following, we partition the interval $[0,2\pi]$ for $\varphi_2$ into four subintervals of size $\pi/2$, defined as
\begin{equation}
    \mathcal{U}_{i} = \left[ \frac{(i-1)\pi}{2},\frac{i\pi}{2}\right] \quad \text{with} 
    \quad i=1,2,3,4. 
    \label{eq:intervals}
\end{equation}

Fig. \ref{fig:t_vs_r_0-1d57}, which is limited to $\varphi_2 \in \mathcal{U}_1$, reveals two distinct behaviors. To clarify them, let us introduce $r_1$ and $r_2$ ($0\le r_1<r_2$) to label the two minima of the AMFPT seen in panels \ref{fig:t_vs_r_0-1d57}a-b. In these panels, when $\varphi_2$ is close to the target location from below, the global minimum is $r^{\ast} =r_2$, while $r_1$ acts as a secondary, metastable minimum. As $\varphi_2$ decreases from the target location, the global minimum abruptly jumps to $r_1$ (which in this case is 0). For sufficiently small $\varphi_2$, the minimum at $r_2$ vanishes completely, leaving the lower resetting rate  as the unique global minimum, i.e., $r^{\ast}=0$.

In the yellow curve of panel \ref{fig:t_vs_r_0-1d57}b, the local minima found at  $r_2\sim 5.8>0$ and at $r_1=0$ have exactly the same AMFPT value. We denote the special value of $\varphi_2$ with these equal minima as $\varphi_{2t}^{(1)}$, and refer to it as the transition value. An arbitrarily small variation around $\varphi_{2t}^{(1)}$ selects one of the two minima as the global one. For example, in panel \ref{fig:t_vs_r_0-1d57}b, for $\varphi_{2}<\varphi_{2t}^{(1)}$ the global minimum is at $r^{\ast }=0$, whereas for $\varphi
_{2}>\varphi_{2t}^{(1)}$ the optimal resetting rate satisfies $r^{\ast }>0$. Therefore, the global minimum of the AMFPT undergoes a discontinuous transition in the optimal parameter $r^{\ast}$, characterized by an abrupt jump $\Delta r=r_{2}-r_{1}>0$, at $\varphi
_{2}=\varphi_{2t}^{(1)}$. Importantly, the two local minima at $r_1$ and $r_2$, and the transition coordinate $\varphi_{2t}$ exist as soon as $m$ is non-zero and below a critical value denoted as $m_{tc}$ (where the subscript $tc$ refers to {\em tri-critical}, see below).

Finally, from panels \ref{fig:t_vs_r_0-1d57}c-d, we observe a continuous decay of the optimal resetting rate $
r^{\ast }$ as $\varphi_{2}$ decreases from the target location, at a fixed $m$ above $m_{tc}$. For instance, when 
$m=2$, the optimal rate decreases continuously from the high value $r^{\ast }\approx
628$ (not visible in the range plotted) in the green curve, to $r^{\ast }=0$ in the black curve. The transition to $r^{\ast }=0$ occurs at $\varphi_{2t}$ and it is continuous. In these cases where $m>m_{tc}$, no abrupt jumps occur.

%%%%%%%%%%%% Figure 5 %%%%%%%%%%%%%%%
%%%%%%%%%%%%%% End of figure 5 %%%%%%

Panel \ref{fig:t_vs_r_0-1d57}c where $m=m_{tc}\approx0.81985\ldots$ represents a special case. This value, associated to the transition angle $\varphi_2=\varphi^{(1)}_{tc} \approx 0.1481\pi$, represents a ``tri-critical" point, which separates $m$-values producing abrupt jumps in $r^{\ast}$ ($m<m_{tc}$) as $\varphi_2$ is varied, from those where $r^{\ast}$ decreases and reaches zero continuously at $\varphi_{2t}$ ($m\ge m_{tc}$). The superscript (1) in $\varphi_{tc}^{(1)}$ indicates that we are referring to the tri-critical coordinate located in the interval $\mathcal{U}_1$. As we mention below, since the AMFPT has mirror symmetry around $\pi/2$, there is also a $\varphi_{tc}^{(2)}$ value that occurs in $\mathcal{U}_2$.

The AMFPT exhibits mirror symmetry around $\varphi_{\ast}=\frac{%
	\pi }{2}$ in the interval $\varphi_{2}\in \left[ 0,\pi \right] $, and
likewise around $\varphi_{2}=\frac{3\pi }{2}$ in the interval $\varphi
_{2}\in \left[ \pi ,2\pi \right] $, due to the term $m \cosh \xi _{r}\left( \pi -\left\vert \varphi
_{\ast }-\varphi_{2}\right\vert \right) $ in the denominator of Eq. %
\eqref{s1-13D}. Hence, for fixed $\varphi_*$ and $\varphi_1$ Eq. \eqref{s1-13D} satisfies

\begin{equation}
\left\langle T_{r} (r,\varphi_{2},m)\right\rangle = 
\begin{cases}
\left\langle T_{r} (r,\pi - \varphi_{2},m)\right\rangle \, \, \text{for} \, \, \varphi_{2} \in \mathcal{U}_1\\
\\
\left\langle T_{r} (r,3\pi - \varphi_{2},m)\right\rangle \, \, \text{for} \, \, \varphi_{2} \in \mathcal{U}_3.
\end{cases}
\label{eq:times}
\end{equation}
Consequently, the first relation implies that the behavior of the AMFPT in $\mathcal{U}_1$ is symmetrical to that of $\mathcal{U}_2$, just as the second relation links ${U}_3$ and $\mathcal{U}_4$. Using this symmetry, we can find the ``mirror'' tri-critical angle in $\mathcal{U}_2$ using
\begin{equation}
    \varphi^{(2)}_{tc} = \pi - \varphi^{(1)}_{tc} \approx 0.8519\pi, \label{u2_tri_coordinate}
\end{equation}
and likewise for all the transition values $\varphi_{2t}$. For instance, the particular situation discussed above---corresponding to $m=0.3$ and $\varphi_{2t}^{(1)} \sim 0.2615\pi$---where two equivalent minima appear in the AMFPT (see Fig. \ref{fig:t_vs_r_0-1d57}-b), also arises within the interval $\mathcal{U}_2$. In this case, the transition occurs at
\begin{equation}
    \varphi_{2t}^{(2)} = \pi-\varphi_{2t}^{(1)}\approx0.7386\pi. \label{u2_tran_coordinate}
\end{equation}
We recall that we use superscripts $(1)$ and $(2)$ to distinguish the transition and tri-critical values of $\varphi_{2}$ occurring in $\mathcal{U}_1$ and $\mathcal{U}_2$, respectively.

\subsection{Computation of the critical coordinates}
Now, to compute the tri-critical point, for example the one that occurs in $\mathcal{U}_1$, $ (\varphi_{tc}^{(1)},m_{tc})$---due to Eq. \eqref{u2_tri_coordinate}, working with interval $\mathcal{U}_1$ is sufficient to deduce $\varphi_{tc}^{(2)}$---we can draw a parallel to the Landau theory of phase transitions \cite{pal2019landau,boyer_optimizing_2024}. Within this framework, a tri-critical point is identified by the simultaneous vanishing of both the first and second order coefficients of the expansion of the quantity to be minimized (i.e., where the first and second derivatives evaluated at the minimum both vanish). In our case, the tri-critical points of the AMFPT can be computed (when they exist) by solving the equation system (see e.g. \cite{pal2019landau})
\begin{equation}
	\lim_{r\to0} \left[\frac{\partial^n \left\langle T_{r}\left( r,\varphi_{2},m\right) \right\rangle}{\partial r^n} \right ]_{\left(
			\varphi_{2},m\right) =\left(
			\varphi^{(1)}_{tc},m_{tc}\right) }=0,\quad n=1,2
	\label{eq:system}
\end{equation}
Due to the form of the AMFPT in Eq. \eqref{s1-13D}, obtaining exact analytical solutions is highly non-trivial. Instead, we employed a numerical scheme using Mathematica to calculate the coordinates of the tri-critical point.

To prepare for the discussion of the subsequent sections, we now outline how the so-called {\em critical} coordinates are obtained. Importantly, we distinguish between critical and tri-critical coordinates by noting that the latter correspond to points separating continuous and discontinuous transitions in the optimal resetting rate $r^{\ast}$. In contrast, the critical point characterizes coordinates that separate discontinuous (first-order) transitions from smooth variations of $r^{\ast}$ (without a transition point). Critical points were previously found in two-point resetting on the semi-infinite line \cite{two_resetting_points} (and also with other distributions \cite{abrupt_2025}), and are therefore expected here. This distinction will become clearer in section \ref{sec:phase_diagram} through the introduction of the $r^{\ast}-\varphi_2$ diagram, also referred as the ``order parameter diagram''.%space of optimal parameters. 

%%%%%%%%%%%% Figure 6 %%%%%%%%%%%%%%%
\begin{figure*}
  \centering
    \includegraphics[width=0.48\linewidth]{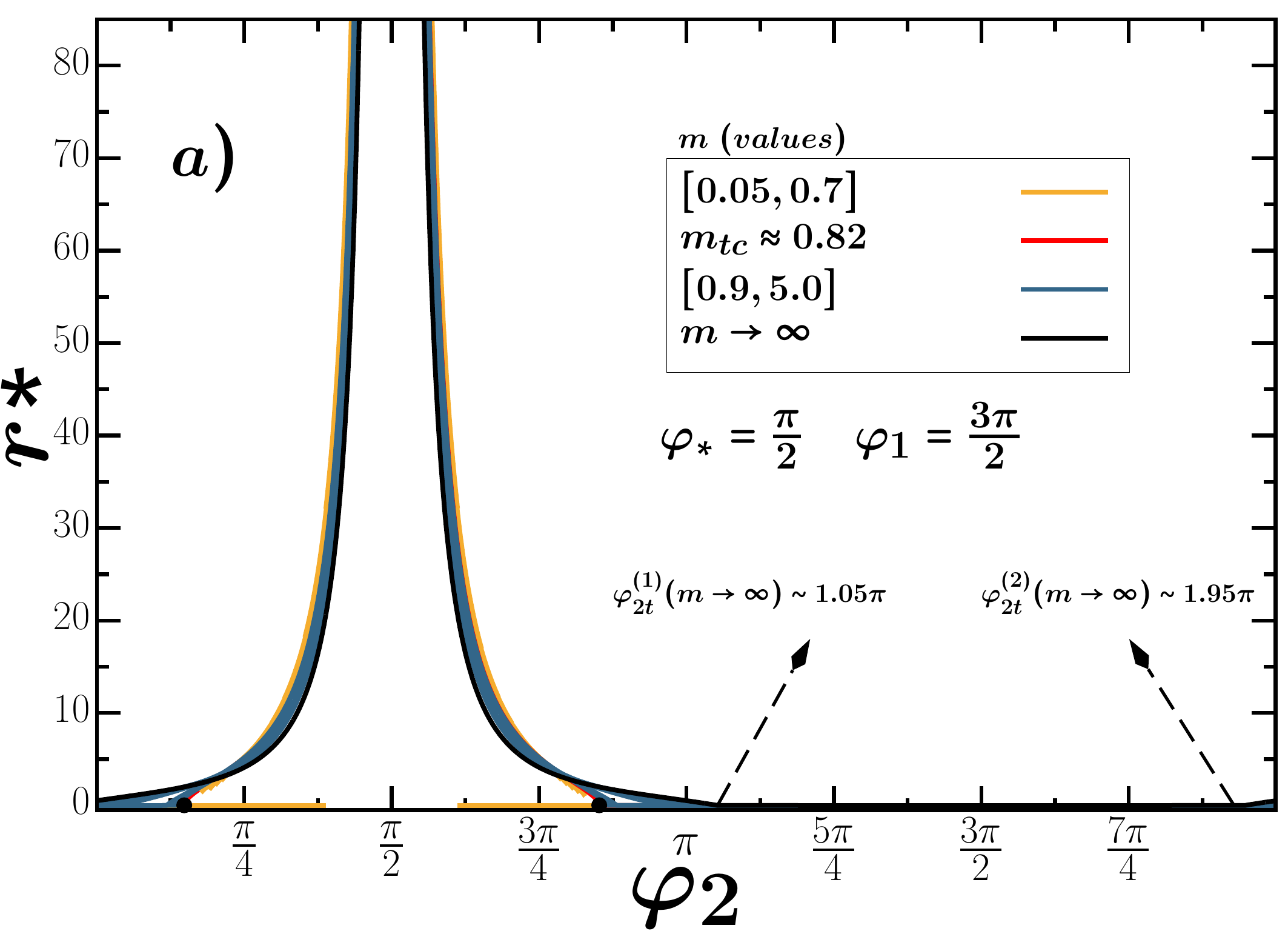}
    \label{fig:all_a}
    \includegraphics[width=0.48\linewidth]{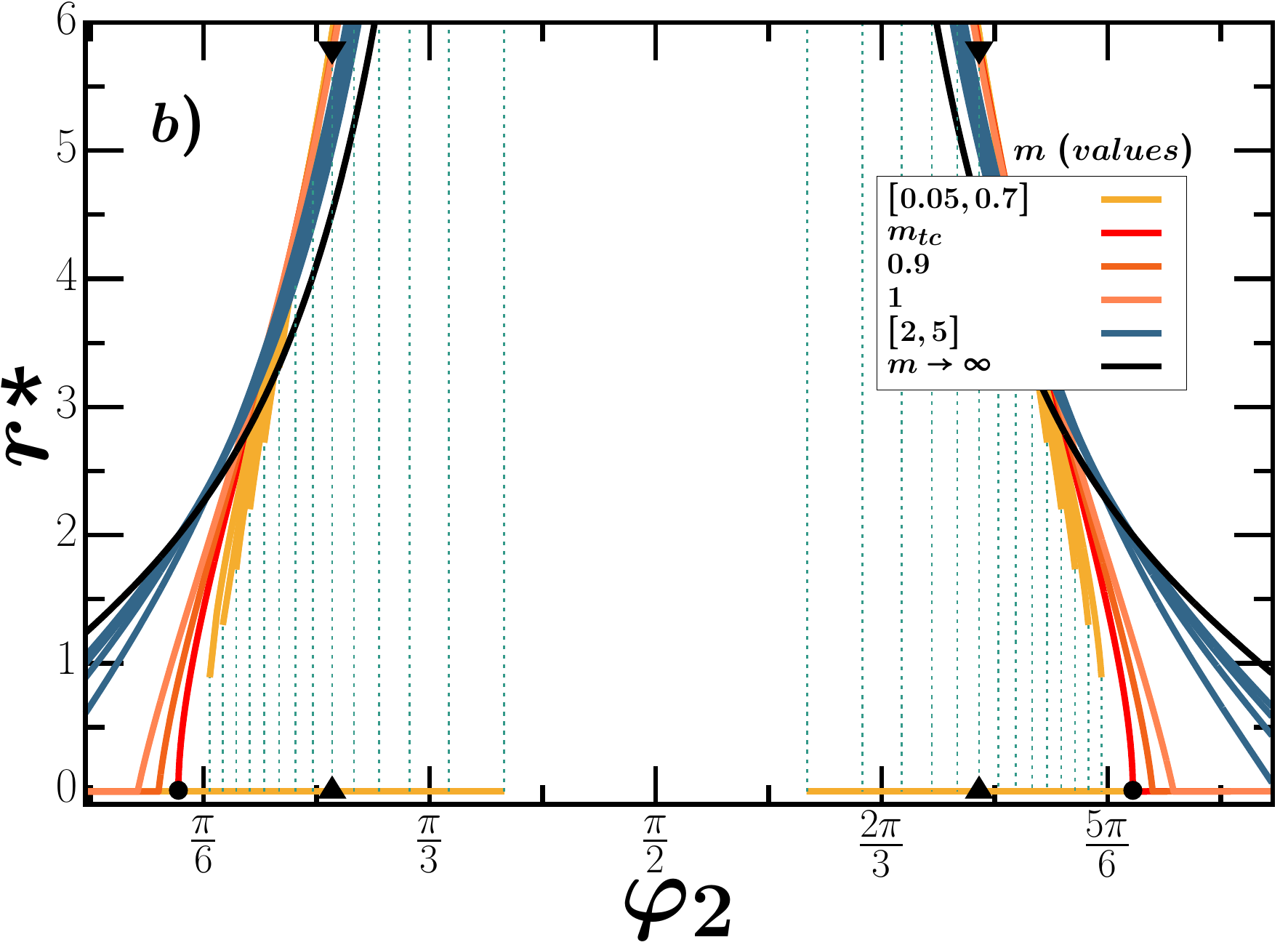}
    \label{fig:u1_b}
  \caption{
Optimal parameter $r^{\ast}$ vs $\protect\varphi_{2}$, defined by
the expression in \eqref{eq:min} for the AMFPT given by Eq. \eqref{s1-13E}, for various values of $m$. Panel b) is a close-up view of panel a) near the tri-critical points (circular black symbols). The dashed vertical lines in panel b) show the discontinuous transitions.  The jump magnitude of $r^{\ast}$ becomes progressively larger as $\varphi_{2t}$ approaches $\pi/2$ ($m$ decreases). See text for other details.} \label{fig:r_vs_phir}
\end{figure*}
%%%%%%%%%%%%%% End of figure 6 %%%%%%

Focusing again on the branch $\mathcal{U}_1$, the critical values  $(\varphi_{2c}^{(1)},m_c)$, if they exist, emerge when $\left \langle T_r \right \rangle$ becomes exceptionally flat and symmetric near its minimum $r^{\ast}$, which takes the value  $r_c$. At $r=r_{c}$, the function not only reaches a minimum--requiring the first derivative to vanish--but also undergoes a transition in concavity, implying a vanishing second derivative. Furthermore, the symmetry of $\left \langle T_r \right \rangle$ on either side of
$r_{c}$ demands the cancellation of any cubic term in the Taylor expansion about $r_c$. These three observations make possible the computations of the critical coordinates, which satisfy (see also \cite{two_resetting_points,abrupt_2025}),
\begin{equation}
	\left.\frac{\partial^n \left \langle T_r(r,\varphi_2,m) \right \rangle}{\partial r^n} \right\vert_{\left(
			r,\varphi_2,m\right) =\left(
			r_{c},\varphi^{
            (1)
            }_{c},m_{c} \right) } =0\quad {\rm for}\ n=1,2,3.
	\label{crit-eqs}
\end{equation}
Eqs. \eqref{eq:system} and \eqref{crit-eqs} also allow to calculate the critical and tri-critical points (if they exist) in the remaining intervals $\mathcal{U}_i$ with $i=2,3,4$.
The values $m_{tc}\approx0.81985\ldots$ and $\varphi^{(1)}_{tc} \approx 0.1481\pi$, see Fig. \ref{fig:t_vs_r_0-1d57}c,  were calculated by solving numerically Eqs. \eqref{eq:system}. No critical point exists for $\varphi_1=3\pi/2$, but this situation will change in the next section dedicated to other values of $\varphi_1$.

Based on these results, we can draw the following conclusions about the behavior of the AMFPT when $\varphi_{2}\in\mathcal{U}_1\bigcup\mathcal{U}_2$ (see Fig. \ref{fig:r_vs_phir}a-b): 
\begin{itemize}
    \item \textit{Discontinuous transitions} occur in the global minimum of the AMFPT, driven by sudden jumps in the optimal resetting rate $r^{\ast}$. For $m<m_{tc}$, and $\varphi_2\in(\varphi_{tc}^{(1)},\varphi_{tc}^{(2)})$, the AFMPT exhibits two competing minima. Consequently, $r^{\ast}$ jumps abruptly from zero to a finite value as $\varphi_2$ increases within $\mathcal{U}_1$, and viceversa as it increases within $\mathcal{U}_2$. The  transition occurs when one of the values, $\varphi_{2} = \varphi_{2t} ^{(i)}$, with $i=1,2$, is reached.
    \item \textit{Continuous transitions} occur in the optimal resetting rate for $m\geq m_{tc}$. In this case, there is no second metastable minimum in the AMFPT. Therefore, the optimal resetting rate, which is 0 for $\varphi_{2}<\varphi_{2t}^{(1)}$ in the interval $\mathcal{U}_1$, continuously increases from zero when $\varphi_{2}>\varphi_{2t}^{(1)}$. By symmetry, this continuous behavior also applies to $\mathcal{U}_2$, where $r^{\ast}$ decreases to zero as $\varphi_2$ moves away from the target location.
\end{itemize}

In the next section, we illustrate these first conclusions by studying the optimal resetting rate $r^{\ast}$ as a function of the resetting angle $\varphi_{2}$. Furthermore, we complement the description by considering the cases $\varphi_2\in[\mathcal{U}_3,\mathcal{U}_4]$.

\subsection{Optimal resetting rate}
Figure \ref{fig:r_vs_phir}
presents the order parameter diagram---a terminology adopted by analogy with magnetic phase transitions---of the optimal resetting rate $r^{\ast}$, defined by
\begin{equation}
    r^{*}(\varphi_{2},m) = \underset{r\ge 0}{\arg \min}\,\left\langle T_{r} (r,\varphi_{2},m)\right\rangle\,, \label{eq:min}
\end{equation}
as a function of the resetting angle $\varphi_{2}$. This diagram exhibits the continuous and discontinuous transitions discussed above. Each curve in Fig. \ref{fig:r_vs_phir} corresponds to a fixed value of $m$.
Due to the complex structure of Eq. \eqref{s1-13E}, the solutions of Eq. \eqref{eq:min} (and thus the points shown in Fig. \ref{fig:r_vs_phir}) were obtained numerically.

Panel b) of Fig. \ref{fig:r_vs_phir} provides a close-up view of panel a) around $\pi/2$ (which corresponds to the target position). In both panels, we have also included the tri-critical points represented, from left to right, by the black
points with coordinates $(\varphi
_{tc}^{(1)},m_{tc}) \approx (0.1481\pi,0.8198) $ and $(\varphi _{tc}^{(2)},m_{tc})\approx(0.8519\pi,0.8198)$, respectively.

From panel \ref{fig:r_vs_phir}a, in the interval 
$\varphi_{2} \in \mathcal{U}_{3} \cup \mathcal{U}_{4}$, 
a global minimum with $r^{\ast} > 0$  is not often observed (for any value of $m$) because the two resetting points are quite far from the target. 
More precisely, if 
$\varphi_{2} \geq \varphi_{2t}^{(1)}(m\to\infty) \approx 1.05\pi$ 
and 
$\varphi_{2} \leq \varphi_{2t}^{(2)}(m\to\infty) \approx 1.95\pi$, 
then $r^{\ast} = 0$ for all $m$. 
This interval is indicated in Fig. \ref{fig:r_vs_phir}a 
by the arrows. 

Thus, the optimal strategy is to not reset, no matter the value of $m$, if $\varphi_2\in[1.05\pi, 1.95\pi]$. 

The symmetry of the AMFPT described in Eq. \eqref{eq:times} can now be directly appreciated in Fig. \ref{fig:r_vs_phir}. Due to this mirror symmetry, once the behavior of $r^{\ast}$ as a function of $\varphi_{2}$ is characterized in the  interval $\mathcal{U}_1$, the corresponding behavior in $\mathcal{U}_2$ follows automatically. Focusing therefore on $\mathcal{U}_1$, the yellow curves  represent the values of $m$ for which discontinuous transitions in $r^{\ast}$ occur, namely for $m<m_{tc}$. As $\varphi_{2}$ increases and crosses its transition value, $r^{\ast}$ jumps from $0$ to a finite value (which can be very large and not always visible on the figure). The special case $m=m_{tc}$ is shown by the red curve. If $m\ge m_{tc}$, the transition to a non-zero resetting rate as $\varphi_{2}$ increases is continuous.

As illustrated in panel \ref{fig:r_vs_phir}b, dashed vertical lines have been added to connect the endpoints of the discontinuous branches of $r^{\ast}$. These guides highlight that, as $\varphi_{2}$ tends to $\pi/2$, the magnitude of the jump grows  larger. The example shown in  panel \ref{fig:t_vs_r_0-1d57}b ($m=0.3$), where two equivalent minima of the AMFPT coexist for the same value $\varphi_{2t} \approx 0.2615\pi$, is represented in panel \ref{fig:r_vs_phir}b by the first pair of black triangles. In this case, at the transition value $\varphi_{2t}^{(1)}$, $r^{\ast}$ jumps from $0$ to $\sim 6$. For smaller values of $m$, $\varphi_{2t}$ keeps approaching $\pi/2$ from the left and the jump in $r^{\ast}$ tends to infinity. On the other hand, in the interval $\mathcal{U}_2$, $r^{\ast}$ decreases continuously as $\varphi_{2}$ increases beyond $\pi/2$ until the second transition angle $\varphi_{2t}^{(2)} = \pi - \varphi_{2t}^{(1)} \approx 0.7386\pi$ is reached. At that point, $r^{\ast}$ drops abruptly to zero, as illustrated by the second pair of black triangles. A similar transition mechanism occurs for other values of $m$ less than $m_{tc}$.
%%%%%%%%%%%% Figure 10 %%%%%%%%%%%%%%%
\begin{figure}
\centering
\includegraphics[width=\linewidth]{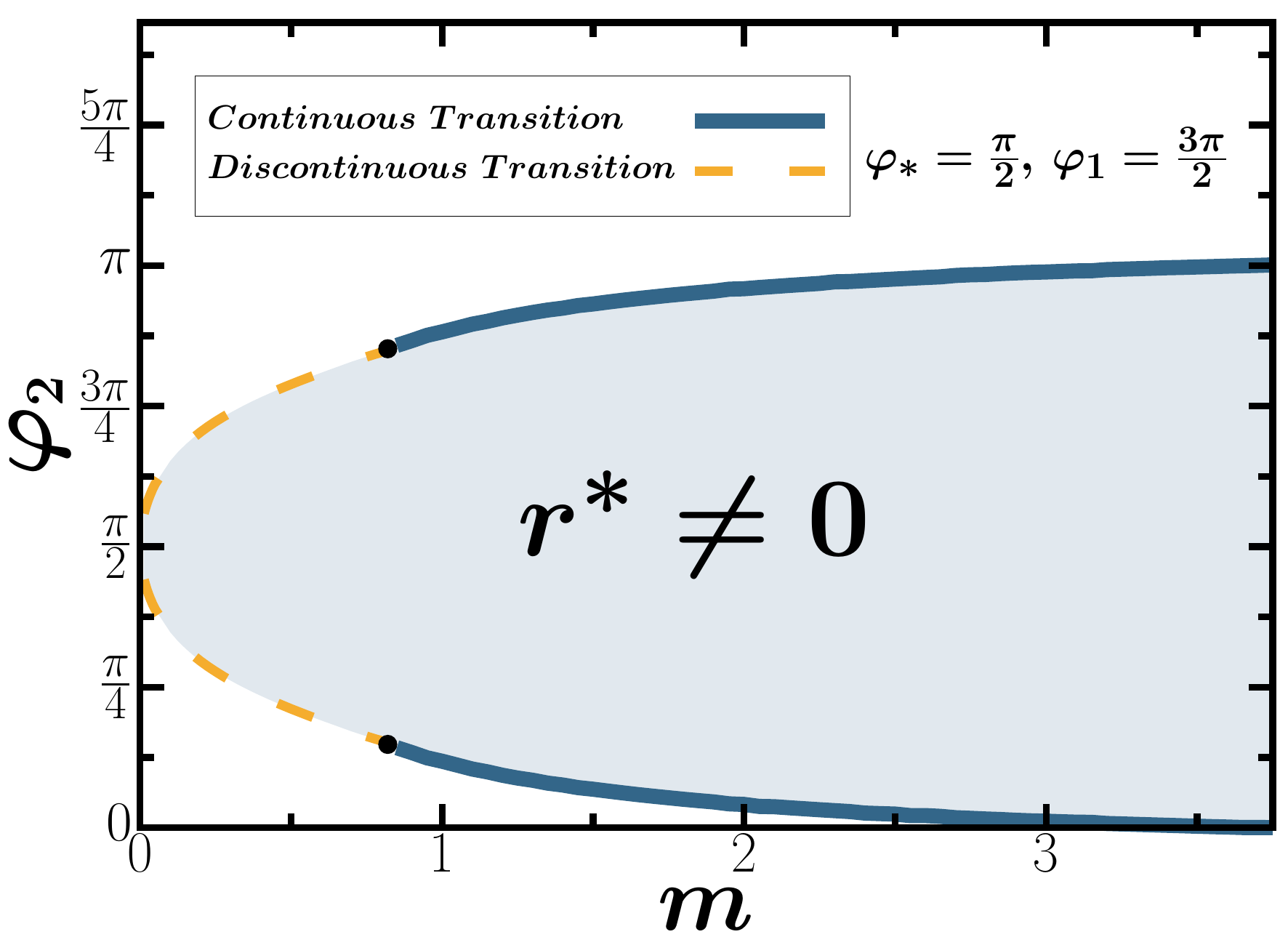}
  \caption{Phase diagram of the case $\varphi_1=3\pi/2$ in the $(m,\varphi_2)$-plane: in the shaded area $r^{\ast}\neq0$, whereas $r^{\ast}=0$ outside. On the solid lines, $r^{\ast}$ reaches $0$ continuously, while each point on the dashed lines represents a discontinuous transition where $r^{\ast}$ abruptly drops to zero.}
	\label{fig:m_vs_phir}
\end{figure}
%%%%%%%%%%%%%% End of figure 10 %%%%%%

\subsection{Phase diagram}\label{sec:phase_diagram}
Up to this point, our discussion has focused on the \lq\lq order parameter" $r^{\ast}$ that minimizes the expression in Eq. \eqref{s1-13D}. 

Now, we summarize the results with $\varphi_1=3\pi/2$ by drawing in Figure \ref{fig:m_vs_phir} a phase diagram in the $(m,\varphi_{2})$-plane. The lines in this plane represent the points where the optimal resetting rate $r^{\ast}$ undergoes a transition to a non-zero value.

The blue-shaded region in the center of Fig. \ref{fig:m_vs_phir} identifies the regime where the optimal resetting rate is non-vanishing ($r^{\ast}\neq0$), whereas the unshaded surrounding region corresponds to $r^{\ast}=0$.

The continuous transitions are shown with solid lines, whereas those that are discontinuous  ($r^{\ast}$ jumps to a finite value) are represented with dashed lines. The points where the solid and dashed segments meet (black dots) mark 
the tri-critical coordinates. 

The curve of Fig. \ref{fig:m_vs_phir} inherits the same symmetry around $\pi/2$ observed in the $r^{\ast}$ versus $\varphi_{2}$ diagram of Fig. \ref{fig:r_vs_phir}. The lower and upper branches in Fig. \ref{fig:m_vs_phir} are composed of all transition points occurring in the interval $\mathcal{U}_1$ and $\mathcal{U}_2$, respectively. The parameter $r^{\ast}$ suffers discontinuous transitions for arbitrarily small weight $m$, provided that the second resetting site $\varphi_2$ is sufficiently near the target. The upper branch increases monotonically with $m$ and asymptotically approaches the non-trivial value $\varphi_{2t}^{(1)}(m\to \infty)\approx 1.05\pi$ shown in Fig. \ref{fig:r_vs_phir}. Similarly, the lower branch reaches the symmetrical value close to $-0.05\pi$.

In the next section, we explore other circular segments in the ring by moving $\varphi_1$ nearer to the target location at $\varphi_{\ast} = \pi/2$.

%%%%%%%%%%%% Figure 7 %%%%%%%%%%%%%%%
\begin{figure}
\centering
\includegraphics[width=\linewidth]{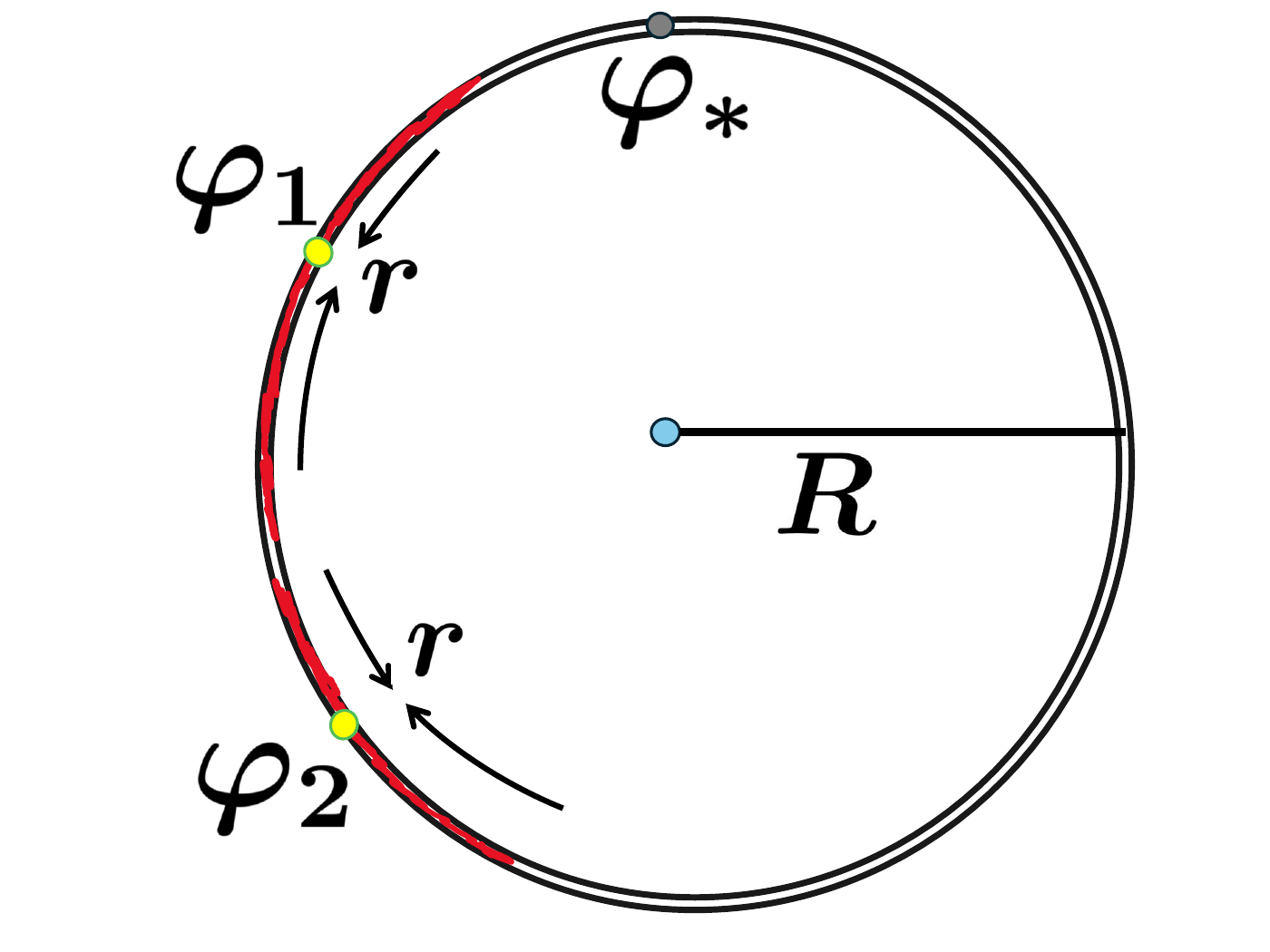}
  \caption{Configuration for other circular segments. Unlike in Fig. \ref{fig:illustration2}, $\varphi_1$ is placed closer to the target site, while we keep considering $\varphi_2$ as a free parameter that varies over the entire interval $[0,2\pi]$. This diagram illustrates the cases $\varphi_1 = 5\pi/6$ and $2\pi/3$ discussed in this work.
}
	\label{fig:illustration3}
\end{figure}
%%%%%%%%%%%%%% End of figure 7 %%%%%%

\section{Other circular segments}\label{other_segments}
In this section, we examine the optimization of the AMFPT  in Eq. \eqref{s1-13D} through the behavior of the optimal restart rate for configurations where the first resetting angle $\varphi_{1}$ (with resetting probability $1/(m+1)$) lies closer to the absorbing target at $\varphi_{\ast} = \pi/2$, compared with the previous case $\varphi_1=3\pi/2$. For illustration, Fig. \ref{fig:illustration3} depicts the set-up considered in the following. A summary of the critical parameters for several values of $\varphi_{1}$ is provided in Table \ref{table}. Due to the symmetry of Eq. \eqref{s1-13D} with respect to $\pi/2$ and $3\pi/2$, only values of $\varphi_1$ in the interval $[\frac{\pi}{2},\frac{3\pi}{2}]$ were considered for constructing Table \ref{table}.
We focus in the following on the representative cases $\varphi_{1} = 5\pi/6$ and $\varphi_{1} = 2\pi/3$.

\subsection{\texorpdfstring{Case $\varphi_1 = \frac{5\pi}{6}$}{Case phi1 = 5pi/6}}

\begin{figure}
\centering
\includegraphics[width=\linewidth]{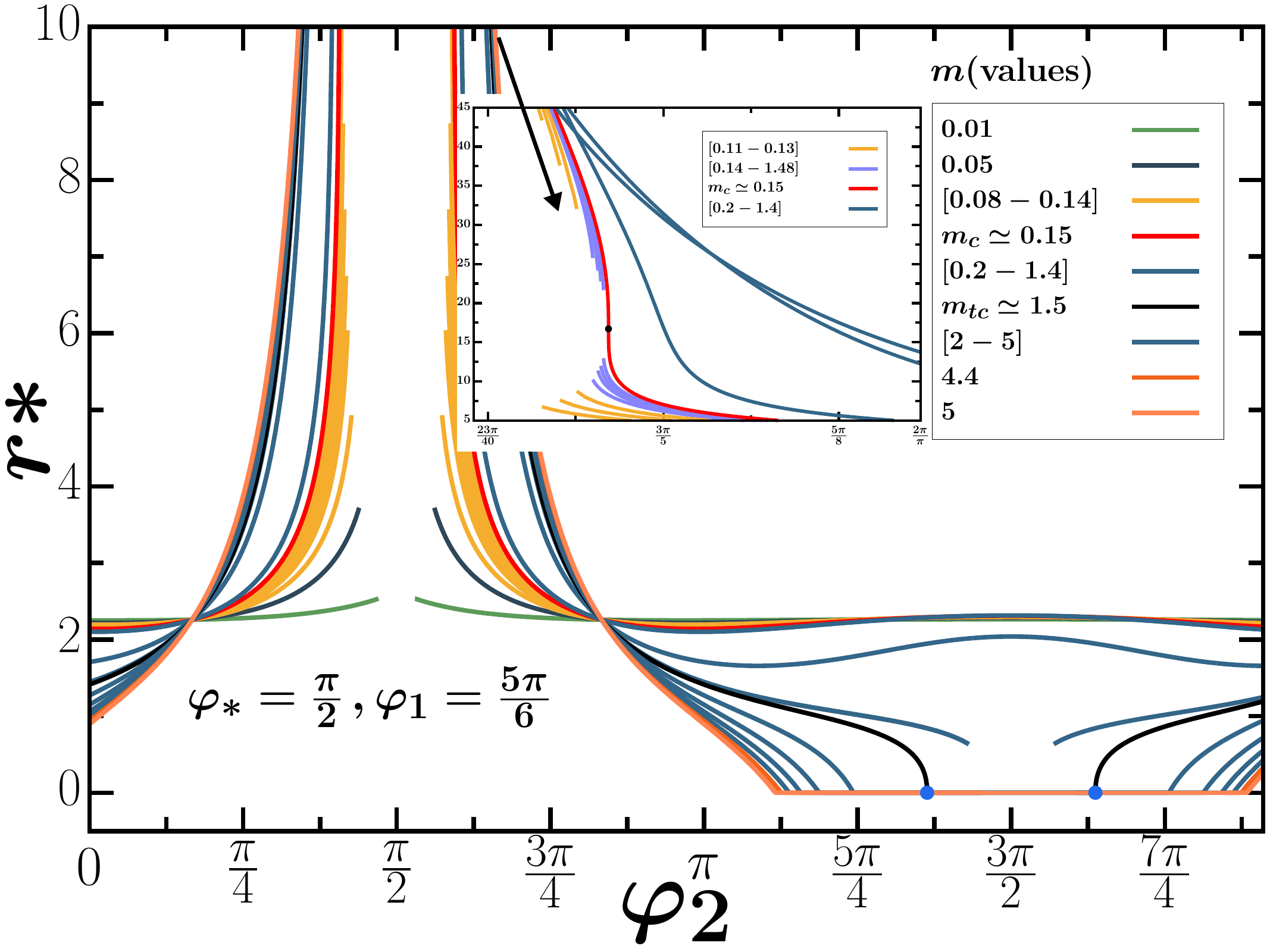}
  \caption{\lq\lq Order parameter" diagram $r^{\ast}$ as a function of $\varphi_2$, for a fixed $\varphi_{1}=5\pi/6$ and several values of $m$. The inset provides a close-up of the transition region to the right of $\pi/2$ indicated by the black arrow, revealing the discontinuous behavior of $r^{\ast}$ near the critical line (in red). The critical point is shown with the black dot. A pair of tri-critical points is marked by blue dots, located symmetrically around $3\pi/2$. These points separate discontinuous and continuous transitions in $r^{\ast}$ as $\varphi_2$ is varied.}
	\label{fig:other_segments1}
\end{figure}

We now analyze the order parameter diagram for $\varphi_1 = 5\pi/6$ (or, equivalently $\varphi_1 = \pi/6$), see Fig. \ref{fig:illustration3}. The variations of $r^{\ast}$ for this configuration are shown in Fig. \ref{fig:other_segments1}, obtained from numerically solving Eq. \eqref{eq:min}.

In Fig. \ref{fig:other_segments1}, a notable difference with the case $\varphi_1 = 3\pi/2$ resides in the fact that the optimal resetting rate $r^{\ast}$ remains nonzero for all $\varphi_{2} \in \mathcal{U}_1\bigcup\mathcal{U}_2$. Intuitively, resetting is more favorable than pure diffusion because the two resetting points are quite close to the target (at $\pi/2$), unlike in the case $\varphi_1 = 3\pi/2$, the furthest possible point along the circular path. 
As described below, the diagram exhibits new features and a higher complexity.

A key difference between the case $\varphi_{1} = 3\pi/2$ is the absence of tri-critical points in the interval $\varphi_{2} \in \mathcal{U}_{1} \bigcup \mathcal{U}_{2}$. However, this region exhibits {\em critical} points that solve Eq. (\ref{eq:system}), with coordinates $(\varphi^{(1)}_{c}, r_{c}, m_{c}) \approx (0.4078\pi, 16.7124, 0.1515)$ and, by symmetry, $(\varphi_{c}^{(2)}=\pi - \varphi^{(1)}_{c}, r_{c}, m_{c}) \approx (0.5922\pi, 16.7124, 0.1515)$. A close-up of the interval $\mathcal{U}_2$ near $(\varphi_{c}^{(2)}, r_{c}, m_{c})$ (inset) reveals a change between smooth and discontinuous behaviors in $r^{\ast}$, separated by a critical line (in red). At the critical point (indicated with a black dot), the derivative of $r^{\ast}$ with respect to $\varphi_2$ is $-\infty$.

By contrast, in the interval $\varphi_{2} \in \mathcal{U}_{3} \bigcup \mathcal{U}_{4}$ in Fig. \ref{fig:other_segments1}, we identify two tri-critical points with parameters $(\varphi^{(1)}_{tc}, m_{tc}) \approx (1.3631\pi, 1.4963)$ and $(\varphi^{(2)}_{tc}=3\pi-\varphi^{(1)}_{tc}, m_{tc}) \approx(1.6369\pi, 1.4963)$, and no critical point. 

%%%%%%%%%%%% Figure 8 %%%%%%%%%%%%%%%
%%%%%%%%%%%%%% End of figure 8 %%%%%%

We first discuss the discontinuities of $r^{\ast}$ found in the interval $\varphi_2\in\mathcal{U}_{1}\bigcup\mathcal{U}_2$. For $m>m_c$, the variations of $r^{\ast}$ with $\varphi_2$ are monotonous and smooth on each side of the divergence at $\pi/2$ (the target position).
For $m<m_c$ (e.g., the yellow curves of Fig. \ref{fig:other_segments1}) the optimal rate is also monotonically increasing in the interval $0\le\varphi_2\le\pi/2$ and diverges at $\pi/2$, but it suffers a discontinuity upwards at a transition value $\varphi_{2} = \varphi_{2t}^{(1)}<\pi/2$. The symmetric variations are observed in the interval $\mathcal{U}_2$ as we increase $\varphi_{2}$ from the target site. 

In the interval $\varphi_{2}\in \mathcal{U}_3\bigcup\mathcal{U}_4$, if $m\le m_c$, $r^{\ast}$ varies smoothly and is nearly flat (without any other first or second order transitions, see the yellow curves of Fig. \ref{fig:other_segments1}). If $m>m_c$ but close to $m_c\approx 0.1515$, a non-monotonous behavior is observed, though. As $m$ further increases and crosses the value $\approx 1.38$, two discontinuous transitions are present, such that $r^{\ast}=0$ in an interval symmetrical around $3\pi/2$. The specific threshold $m=1.38$ characterizing the onset of these first-order transitions is non-trivial and is computed via numerical methods.
The discontinuous transitions disappears at $m= m_{tc}=1.4963$, the tri-critical parameter mentioned above. The tri-critical points are represented by the blue dots in Fig. \ref{fig:other_segments1}, with coordinates $(\varphi^{(1)}_{tc},m_{tc})$ 
and $(\varphi_{tc}^{(2)},m_{tc})$. 
For $m\ge m_{tc}$, the optimal rate undergoes two continuous transitions in the interval $\varphi_{2}\in \mathcal{U}_3\bigcup\mathcal{U}_4$. 

Furthermore, Fig. \ref{fig:other_segments1} exhibits a behavior already observed in Fig. \ref{fig:r_vs_phir}, when $\varphi_1=\varphi_2$. Here, two points in Fig. \ref{fig:other_segments1}, given by $(\pi/6,2.2592)$ and $(5\pi/6,2.2592)$, belong to all the curves, independent of the value of $m$. This occurs because $\varphi_2=\pi/6$ and $\varphi_2=5\pi/6$ are symmetrically located with respect to $\pi/2$ and these cases are equivalent to having a single resetting point.

In the following section, we explore how this behavior of $r^{\ast}$ evolves when $\varphi_1$ is placed even closer to the target.

\subsection{\texorpdfstring{Case $\varphi_1 = \frac{2\pi}{3}$}{Case phi1 = 2pi/3}}

\begin{figure}
\centering
\includegraphics[width=\linewidth]{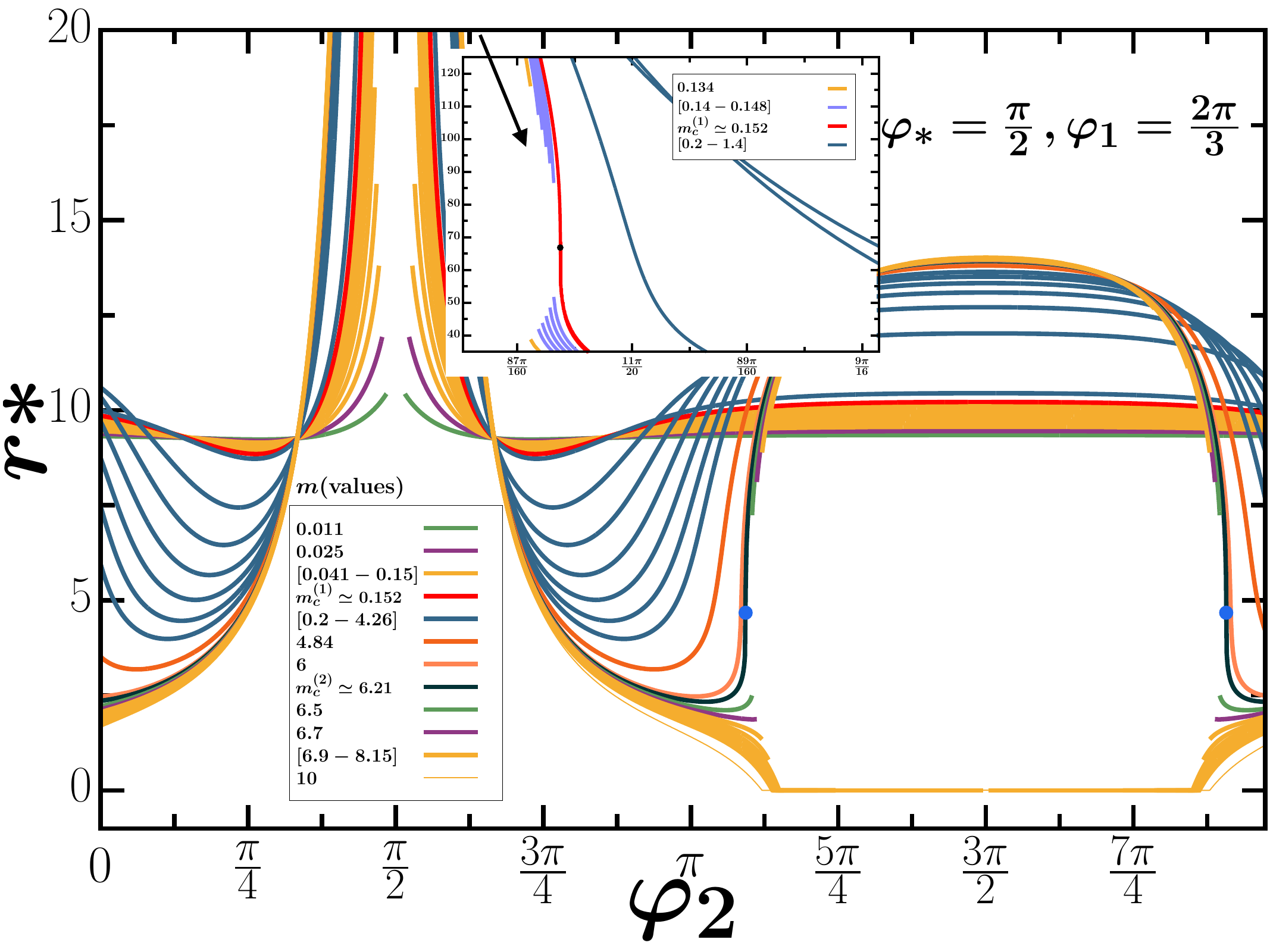}
  \caption{Optimal resetting rate as a function of ${\varphi_2}$ for the case $\varphi_1 = 2\pi/3$ and several values of $m$. As in the previous case $\varphi_1 = 5\pi/6$, this configuration exhibits a rich phenomenology. The inset provides a magnified view of the region indicated by the black arrow in the main panel, with the critical point highlighted in black. A second pair of critical points, located symmetrically around $3\pi/2$, is indicated by blue dots.
}
	\label{fig:other_segments2}
\end{figure}

When the first resetting site is placed even closer to the target, at $\varphi_1 = \frac{2\pi}{3}$ (or equivalently $\pi/3$), the optimal parameter diagram also exhibits qualitative differences from the two previous cases, see Fig. \ref{fig:other_segments2}. 
Notably, tri-critical points are absent in the full interval $\varphi_2\in[0,2\pi]$. Instead, we identify two  pairs of critical coordinates that characterize first-order transitions in the optimal resetting rate $r^{\ast}$.

The first pair of critical points is similar to the one of Fig. \ref{fig:other_segments1} and has critical coordinates $(\varphi_{c}^{(1)},r_c,m^{(1)}_c) \approx (0.4539\pi,66.85,0.1515)$ in $\mathcal{U}_1$, and its mirror counterpart $(\varphi_{c}^{(2)},r_c,m^{(1)}_c) \approx (0.5461\pi,66.85,0.1515)$ in $\mathcal{U}_2$. The inset of the figure is a close-up that displays again the discontinuities in $r^{\ast}$ that emerge from the critical line (in red).

The second pair of critical parameters, indicated by the blue dots in the main panel of Fig. \ref{fig:other_segments2}, are given by $(\varphi_{c}^{(3)},r_c,m^{(2)}_c) \approx (1.0936\pi,4.6744,6.2094)$ and its mirror counterpart $(\varphi_{c}^{(4)},r_c,m^{(2)}_c)\approx(1.9071\pi,4.6744,6.2094)$. The superscripts in $\varphi^{(i)}_c$, with $i=1,2,3,4$, indicate the associated interval $\mathcal{U}_i$.

%%%%%%%%%%%% Figure 9 %%%%%%%%%%%%%%%
%%%%%%%%%%%%%% End of figure 9 %%%%%%

%%%%%%%%%%%% Table 1 %%%%%%%%%%%%%%%%%%%%
\begin{table*}[t]
    \begin{ruledtabular}
    \begin{tabular}{ccc} % No vertical lines (|) allowed in PRE style
        \multicolumn{3}{c}{Critical and Tri-critical Coordinates}\\
        \colrule
        First Resetting Angle $\varphi_1$ & Tri-Critical $(\varphi^{(i)}_{tc},m_{tc})$ & Critical $(\varphi^{(i)}_c,m_c,r_{c})$ \\
        \colrule
        $2\pi/3$ & N/A & $(0.4539\pi,\,0.1515,\,66.8506) \land (1.093\pi,\,6.2094,\,4.6744)$ \\
$5\pi/6$ & $(1.3631\pi,\,1.4964)$ & $(0.4078\pi,\,0.1515,\,16.7124)$ \\
$\pi$ & N/A & $(0.3615\pi,\,0.1525,\,7.3145)$ \\
$7\pi/6$ & N/A & $(0.3025\pi,\,0.1957,\,2.4401)$ \\
$4\pi/3$ & $(0.1769\pi,\,0.6006)$ & N/A \\
$3\pi/2$ & $(0.1481\pi,\,0.8199)$ & N/A \\
    \end{tabular}
    \end{ruledtabular}
    \caption{The AMFPT in Eq. \eqref{s1-13D} exhibits a level of complexity that allows critical and tri-critical points to exist in the order parameter diagram. This table summarizes critical and tri-critical values found for different angular values of $\varphi_1$. The superscript (i) in the second and third columns denotes the specific interval $\mathcal{U}_i$ in which the corresponding angular coordinate $(\varphi^{(i)}_{tc}$ or $\varphi^{(i)}_{c})$ resides. N/A is a shortcut to not applicable.}
    \label{table}
\end{table*}

%%%%%%%%%%%%% End Table 1 %%%%%%%%%%

Rather than detailing the entire angular domain, we choose to characterize the system through its differences with the configurations previously studied. Near the target, specifically in the interval $\varphi_2 \in [\pi/3,2\pi/3]$, the behavior remains qualitatively similar to the case $\varphi_1=5\pi/6$, with larger discontinuous jumps as $\varphi_2$ approaches $\frac{\pi}{2}$ for $m<m^{(1)}_c$. However, when $\varphi_{2}\in [0,\pi/3]$, the optimal resetting rate $r^{\ast}$ exhibits a new non-monotonic dependence on $\varphi_2$, a feature that disappears for values of $m$ greater than $\approx 6$ (orange curve), above which $r^{\ast}$ monotonically increases with $\varphi_2$. Furthermore, analogous to the previous cases, there are two symmetrical points located around $\pi/2$ where all the curves intersect regardless of the value of $m$, specifically at $(\pi/3,9.2635)$ and $(2\pi/3,9.2635)$. Clearly, these angles correspond to configurations that are equivalent to a single-resetting point located at $\varphi_1$.

A rather complex phenomenology arises in the region $\varphi_2\in [\pi,3\pi/2]$ (and its symmetrical $[3\pi/2,2\pi]$), where the order parameter diagram exhibits features qualitatively distinct from the configurations previously examined. For $m \gtrsim m_{c}^{(2)}$, $r^{\ast}$ is non-monotonic and undergoes an abrupt jump from a lower finite value $r_1$ to a larger value $r_2$, at a transition coordinate $\varphi_{2t}^{(3)}$. As $m$ increases further, $r^{\ast}$ enters into a monotonic decrease behavior prior to the upward jump, with the lower value of the discontinuity, $r_1$, gradually going to zero. Actually, for $m\ge7.3$, the optimal rate undergoes a second-order transition to 
$r^{\ast}=0$, followed by a discontinuous jump from $r_1 =0$ to $r_2>0$. As $m$ increases further, the interval of angles $\varphi_2$ such that 
$r^{\ast}=r_2\neq0$ shrinks around $3\pi/2$. This regime culminates at $m\approx8.15$, where $\varphi_{2t}^{(3)}\approx 1.49654\pi$ (and where $r^{\ast}$ jumps from $0$ to $r_2\approx14.017$). Beyond this threshold ($m>8.15$), the discontinuous transitions disappear, and the second-order transition to $r_1=0$ remains. 

\section{Conclusions}\label{conclusions}
Diffusion with stochastic resetting in bounded domains is known to exhibit continuous transitions for the optimal resetting rate, which vanishes if the resetting point is sufficiently close to an absorbing boundary \cite{pal2019first}. Here, we focused on the optimization of the averaged MFPT to a target site of a Brownian particle confined to a ring and subject to resetting to two positions. This problem exhibits a rich phenomenology of phase transitions. We characterized the behavior of the optimal restart rate, $r^{\ast}$, in terms of the parameters of the system, i.e., the resetting positions/angles and the resetting weight of each position (defined through the parameter $m$). We developed such a characterization for two different types of configurations: i) as a preliminary, one consisting of a single resetting site, at position $\varphi_1$, different from the initial angle $\varphi_0$ of the particle at $t=0$, and ii) resetting to a pair of sites, in which case the initial angle is randomly chosen to be one of the resetting sites with its corresponding weight. As a first step, we recalled the renewal theory of diffusion with stochastic resetting to a distribution of points, which involves the survival probability of the particle without resetting.

For the single resetting site arrangement, we applied a modified CV condition--a criterion that, if satisfied, predicts the decrease of the mean search time thanks to resetting--and drew a phase diagram. Given a pair $(\varphi_0,\varphi_1)$, this diagram specifies whether the resetting mechanism expedites the search of a target located at $\varphi_{\ast}=0$ or not, see, Fig. \ref{fig:dia_phir_vs_phi0}. When $\varphi_1 \neq \varphi_0$, choosing $\varphi_0 =\pi$ maximizes the interval of angles $\varphi_1$ for which resetting is advantageous. In contrast, if one fixes $\varphi_0=\varphi_1$, the interval of angles where resetting is beneficial has a smaller size. 

In the second part of this work we examined the averaged mean first-passage time (AMFPT) of the two-resetting site set-up, given by  Eq. \eqref{s1-13D}. With the aid of a numerical scheme we characterized the optimal resetting rate $r^{\ast}$ as a function of the second resetting angle and its relative weight $m$, for representative values of the first resetting site $\varphi_1$. The AMFPT admits useful analytical expressions in some limiting regimes ($m\to0$, $m\to \infty$ and $r\to 0$, respectively) and displays mirror symmetries that reduce the number of configurations to be studied.

The \lq\lq order parameter" diagrams in Figs. \ref{fig:r_vs_phir}, \ref{fig:other_segments1} and \ref{fig:other_segments2}, obtained for representative values of $\varphi_1$ constitute one of the main results of this study. These diagrams show a rich mixture of second-order transitions, where $r^{\ast}$ continuously reaches zero as $\varphi_2$ is varied, and discontinuous transitions, characterized by abrupt jumps between $r^{\ast} =0$ and $r^{\ast}>0$, or two non-zero values. Fixing $\varphi_1$, notable outcomes include the existence of both critical and tri-critical points, observed for $\varphi_1 =5\pi/6$ in Fig. \ref{fig:other_segments1}, and the existence of two pairs of critical points, shown in Fig. \ref{fig:other_segments2}. For $\varphi_1 = 3\pi/2$ we identified an interval (approximately $[1.05\pi,1.94\pi]$ shown in Fig. \ref{fig:r_vs_phir}) where $r^{\ast} =0$ for all $m$. The critical and tri-critical coordinates for the representative values of $\varphi_1$ considered in this study were listed in Table \ref{table}. Each value of $\varphi_1$ studied yields peculiar first and second-order transitions, where the dependence of $r^{\ast}$ on $\varphi_2$ and $m$ requires a separate analysis. We also draw the phase diagram in the $(m,\varphi_{2})$-plane in Fig. \ref{fig:m_vs_phir} for the case where $\varphi$ and the target site were diametrically opposite. In this representation, solid and dashed segments, respectively, indicate the continuous and discontinuous transitions in the global minimum of the AMFPT, while their intersections (black symbols) define the tri-critical points. 

Together, these results demonstrate that the diagrams for the optimal resetting rate change dramatically as the first resetting site $\varphi_1$ approaches the target site, located at $\varphi_{\ast}$. The locations of the resetting sites respective to the target, as well as their weights, therefore play a decisive role in the optimization of diffusion with stochastic resetting to a pair of sites. Finally, the transitions found here provide a comprehensive foundation for future analytical refinements and for exploring other resetting distributions.

\vspace{1cm}
\noindent {\bf{Acknowledgments}} P.J. acknowledges financial support from SECIHTI via scholarship No. 4018217 for Ph.D. during the course of this research. D.B. acknowledges support from Conacyt (SECIHTI), grant Ciencia de Frontera 2019 no. 10872. P.J., P.C. and L.D.  acknowledge financial support from the Frontiers Science grant No. CBF-2025-I-3987, financed by SECIHTI. \\

\appendix

\section{Free propagator of a Brownian particle in a ring} \label{surv_eq}
On general grounds, we can use the Markov property of normal diffusion to write
\begin{equation}
	G_0\left( \varphi,t|\varphi_{0}\right) =\int_{0}^{t}\mathcal{P}\left( \varphi,\tau |\varphi_{0}\right)
	G_0\left( \varphi,t-\tau |\varphi\right) \dd\tau ,  \label{eq:free_prop}
\end{equation}
where $G\left( \varphi,t|\varphi_{0}\right)$ is the probability density function (in the absence of absorbing boundaries) for finding the particle in position $\varphi$ at time $t$ if it started at $\varphi_0$ at $t=0$, and  $\mathcal{P}\left( \varphi,\tau |\varphi_{0}\right)$ is the PDF of the first-passage time $\tau$ at $\varphi$ with the same initial condition. Laplace transforming this equation gives us
\begin{equation*}
	\tilde{G}_{0}\left( \varphi,s|\varphi_{0}\right) =\mathcal{\tilde{P}}\left(
	\varphi,s|\varphi_{0}\right) \tilde{G}_{0}\left( \varphi,s|\varphi\right),
\end{equation*}
and solving for $\mathcal{\tilde{P}}\left( \varphi,s|\varphi_{0}\right) $, we obtain
\begin{equation}
	\mathcal{\tilde{P}}\left( \varphi,s|\varphi_{0}\right) =\frac{\tilde{G}_{0}\left(
		\varphi,s|\varphi_{0}\right) }{\tilde{G}_{0}\left( \varphi,s|\varphi\right) }.  \label{lap_free_prop}
\end{equation}
Additionally, from the first passage theory we know that, if we have an
absorbing target at $\varphi=0$ then the survival probability $Q\left(
\varphi_{0},t\right) $ is related to $\mathcal{P}\left( \varphi,t|\varphi_{0}\right) $ via the
probability of reaching $x=0$ at some time $t+\dd t$: $\mathcal{P}\left(
0,t|\varphi_{0}\right) \dd t=Q\left( \varphi_{0},t\right) -Q\left( \varphi_{0},t+\dd t\right) $. Hence, the relation
\begin{equation}
	\mathcal{P}\left( 0,t|\varphi_{0}\right) =-\frac{\partial Q\left( \varphi_{0},t\right) }{
		\partial t},  \label{time-prop}
\end{equation}
and its Laplace transform is straightforward $\mathcal{\tilde{P}}\left(
0,s|\varphi_{0}\right) =1-s\tilde{Q}_{0}\left( \varphi_{0},t\right) $, or
\begin{equation*}
	\tilde{Q}_{0}\left( \varphi_{0},t\right) =\frac{1-\mathcal{\tilde{P}}\left(
		0,s|\varphi_{0}\right) }{s}.
\end{equation*}
If we substitute Eq. \eqref{lap_free_prop} in this last equation, we obtain
\begin{equation}
	\tilde{Q}_{0}\left( \varphi_{0},s\right) =\frac{1}{s}\left( 1-\frac{\tilde{G}
		_{0}\left( \varphi=0,s|\varphi_{0}\right) }{\tilde{G}_{0}\left( \varphi=0,s|\varphi\right) }\right) .
	\label{lap_trans}
    \end{equation}
We now calculate $G_{0}\left(
\varphi ,t|\varphi _{0}\right) $ for a Brownian particle diffusing freely in
the circle with initial position localized at the coordinates $\left(
R,\varphi _{0}\right) $, where $R$ is the radius of the circle. The forward
Fokker-Planck equation for this problem can be written in polar coordinates
as%
\begin{equation}
	\frac{\partial G_{0}\left( \varphi ,t|\varphi _{0}\right) }{\partial t}=%
	\frac{D}{R^{2}}\frac{\partial ^{2}G_{0}\left( \varphi ,t|\varphi _{0}\right) 
	}{\partial \varphi ^{2}},  \label{A1-15}
\end{equation}%
in which $\varphi \in \left[ 0,2\pi \right] $. If we propose the solution $%
G_{0}\left( \varphi ,t|\varphi _{0}\right) =A\left( \varphi \right) B\left(
t\right) $, then, upon substitution in Eq. \eqref{A1-15} we obtain the following two equations (with $\lambda$ a constant)
\begin{equation}
	\frac{\partial B\left( t\right) }{\partial t}=-\lambda  \label{s1-16}
\end{equation}%
and%
\begin{equation}
	\frac{D}{R^{2}}\frac{\partial ^{2}A\left( \varphi \right) }{\partial \varphi
		^{2}}=-\lambda .  \label{s1-17}
\end{equation}%
The solution of \eqref{s1-16} is simply $B\left( t\right) =B\left(
0\right) e^{-\lambda t}$, with $B\left( 0\right) $ a constant to be
determined later. Furthermore, the solution of \eqref{s1-17} is
\begin{equation*}
	A\left( \varphi \right) =C\cos \left( \kappa \varphi \right) +D\sin \left(
	\kappa \varphi \right) ,
\end{equation*}%
with $\kappa ^{2}=\frac{R^{2}\lambda }{D}$. Since $A\left( \varphi \right) $
must be invariant under rotations, $A\left( \varphi \right) $ must satisfy the
condition $A\left( \varphi +2\pi \right) =A\left( \varphi \right) $. This condition is satisfied if $\kappa =m=1,2,3...$ from which we obtain%
\begin{equation*}
	A\left( \varphi \right) =C\cos \left( m\varphi \right) +D\sin \left(
	m\varphi \right) .
\end{equation*}%
Therefore, the final solution can be written as%
\begin{equation}
	G_{0}\left( \varphi ,t|\varphi _{0}\right) =\sum_{m=0}^{\infty }\{ C\cos
	\left( m\varphi \right) +D\sin \left( m\varphi \right) \} e^{-\frac{%
			m^{2}Dt}{R^{2}}},
	\label{s1-18}
\end{equation}%
with
\begin{equation}
    \lambda =\frac{m^{2}D}{R^{2}}.\label{s1-18b}
\end{equation}
The constants $C$ and $D$ can be computed using the initial condition $G_{0}\left( \varphi ,t=0|\varphi
_{0}\right) =\delta \left( \varphi -\varphi _{0}\right) $, which upon application we find
\begin{equation*}
	C_{0}=\frac{1}{2\pi},\text{ and }C=\frac{\cos m\varphi _{0}}{\pi}\text{, 
	}D=\frac{\sin m\varphi _{0}}{\pi}.
\end{equation*}%
Finally, by direct substitution into Eq. \eqref{s1-18} and further simplifications, we find the required propagator%
\begin{equation}
	G_{0}\left( \varphi ,t|\varphi _{0}\right) =\frac{1}{2\pi}+\frac{1}{\pi }%
	\sum_{m=1}^{\infty }e^{-\frac{m^{2}}{R^{2}}Dt}\cos m\left( \varphi -\varphi
	_{0}\right) .  \label{s1-19}
\end{equation}
Note that the corresponding propagator for the arc-length coordinate, $y=R\varphi$ with $y \in [0,2\pi R]$, is obtained by rescaling the angular propagator. Because $\dd y = R\dd \varphi$, the probability density must acquire the reciprocal factor $R$, this is to remain properly normalized.

The Laplace transform of Eq. \eqref{s1-19} is:
\begin{equation}
	\tilde{G}_{0}\left( \varphi ,s|\varphi _{0}\right) =\frac{1}{2\pi}\left( 
	\frac{1}{s}+\frac{2R^{2}}{D}\sum_{m=1}^{\infty }\frac{\cos m\left( \varphi
		-\varphi _{0}\right) }{m^{2}+\frac{sR^{2}}{D}}\right) \label{s1-20}
\end{equation}
and using line 1.445-2 of \cite{gradshteyn2007table} results in
\begin{equation}
	\sum_{m=1}^{\infty }\frac{\cos mx}{m^{2}+a^{2}}=\frac{\pi }{2a}\frac{\cosh 
		\left[ a\left( \pi -x\right) \right] }{\sinh \left( a\pi \right) }-\frac{1}{
		2a^{2}}\text{ \ }x\in \left[ 0,2\pi \right] \label{series_relation}
\end{equation}
Eq. \eqref{s1-20} may be rewritten as
\begin{equation}
	\tilde{G}_{0}\left( \varphi ,s|\varphi _{0}\right) =\frac{R}{\sqrt{4sD}} 
	\frac{\cosh \left[ \xi_s \left( \pi -\left\vert \varphi -\varphi
		_{0}\right\vert \right) \right] }{\sinh \left( \xi_s \pi \right) },  \label{final_exp_prop}
\end{equation}
with $\xi_s =R\sqrt{\frac{s}{D}}$. The direct substitution of this final expression for $\tilde{G}_{0}\left( \varphi ,s|\varphi _{0}\right)$ into Eq. \eqref{lap_trans} gives us
\begin{equation}
    \tilde{Q}_{0}\left(\varphi _{0},s\right) =\frac{1}{r}\left[ 1-\frac{%
		\cosh \xi_{s}\left( \pi -\left\vert \varphi _{\ast }-\varphi
		_{0}\right\vert \right) }{\cosh \xi_{s}\pi }\right], \label{surv_apA}
\end{equation}
which is Eq. \eqref{surv_expl} evaluated at $s$ intead of $r$.

\section{CV criterion} \label{ap:cv}
In this section, we draw attention to Eqs. \eqref{cond_res}-\eqref{ring_mfpt}, and explain how they are used to derive the inequality in Eq. \eqref{cond_res1} for resetting to a single point. These equations help us to explore regimes in the parameter space of the system in which the resetting mechanism makes the MFPT decrease compared to the system without resetting. As the underlying process has not been specified, these expressions are valid not only for Brownian dynamics but also for any other process under consideration; see, for example, \cite{reuveni_optimal_2016} for more details.

The inequality in \eqref{cond_res1} can be obtained from \eqref{cond_res} by expanding the Laplace transform of the survival probability $\tilde{Q}_0$ at small $r$
\begin{equation}
    \tilde{Q}_0(\varphi_{i},r) = \int_{0}^{\infty} e^{-rt} Q_0(\varphi_i,t)\, \dd t \quad i=0,1 \label{ap_b1}
\end{equation}
this is
\begin{equation}
	\tilde{Q}_{0}\left( \varphi_{i},r\right) =\int_{0}^{\infty }dt\left( 1-rt+\frac{r^{2}t^{2}}{2}
	-...\right) Q_0(\varphi_{i},t). \label{ap_b2}
\end{equation}%
Denoting $t_0(\varphi_i)$ as the first-passage time of the process without resetting starting from $\varphi_i$,
the MFPT is $T_{0}(\varphi_i) =\langle t_0(\varphi_i)\rangle= \int_{0}^{\infty }dt\, Q_0(\varphi_{i},t)$ and the second moment is
\begin{multline}
	\langle t_{0}^{2}(\varphi_{i})\rangle =-\int_{0}^{\infty }dt\, t^{2}\frac{
		\partial Q_0(\varphi_{i},t) }{\partial t}=\left[ -t^{2}Q_0(\varphi_{i},t) \right] _{0}^{\infty }\\+2\int_{0}^{\infty }dt\, tQ_0(\varphi_{i},t). \label{ap_b3}
\end{multline}
Provided $Q_0(\varphi_{i},t)$ decays to 0 sufficiently fast as $t\to \infty$, it follows that 
\begin{equation}
   \langle t_{0}^{2}(\varphi_{i})\rangle =2\int_{0}^{\infty
	}dt\, tQ_0(\varphi_{i},t). \label{ap_b3b}
\end{equation}
With these results, the Laplace transform of the survival probability can be rewritten at order ${\cal O}(r)$ as
\begin{equation}
    \tilde{Q}_0(\varphi_i,r) = \langle t_{0}(\varphi_{i})\rangle - \frac{r}{2}\langle t_{0}^{2}(\varphi_{i})\rangle +...  \label{ap_b5}
\end{equation}
Consequently, Eq. \eqref{ring_mfpt} can be written as
\begin{equation}
    T_r(\varphi_0) = \frac{\langle t_{0}(\varphi_{0})\rangle - \frac{r}{2}\left\langle t_{0}^{2}(\varphi_{0})\right\rangle +...}{1-r\left( \langle t_{0}(\varphi_{1})\rangle - \frac{r}{2}\left\langle t_{0}^{2}(\varphi_{1})\right\rangle +...\right)}. \label{ap_b6}
\end{equation}
which gives at leading order in $r$,
\begin{equation}
  T_r(\varphi_0) = \langle t_{0}(\varphi_{0})\rangle
  +r\left[ \langle t_{0}(\varphi_{1})\rangle
  \langle t_{0}(\varphi_{0})\rangle
  -\frac{1}{2}\langle t^{2}_0(\varphi_0) \rangle\right]+ \mathcal{O}[r^{2}]. \label{ap_b7_b}
\end{equation}
It is straightforward to see that
\begin{equation}
    \lim_{r\to0} \frac{\dd T(\varphi_0)}{\dd r} =\langle t_{0}(\varphi_{1})\rangle
  \langle t_{0}(\varphi_{0})\rangle
  -\frac{1}{2}\langle t^{2}_0(\varphi_0) \rangle. \label{ap_b9}
\end{equation}
Substituting this last expression into the condition \eqref{cond_res} is how we obtain equation \eqref{cond_res1}.
\bibliography{biblio}
\end{document}